\documentclass[twocolumn,twocolappendix]{aastex631}
\usepackage{times, amsmath, amssymb, amsfonts, latexsym}\input{fleqn.clo}
\usepackage{multirow, graphicx, color, natbib}
\usepackage{hyperref, array, mathtools, enumitem}
\usepackage{soul}
\newcommand{\uat}[2]{\href{http://astrothesaurus.org/uat/#2}{#1 (#2)}}

\def\mpch{\,{h^{-1} {\rm Mpc}}}            
\shorttitle{Wavelet analysis of matter clustering}
\shortauthors{Wang et al.}
\begin{document}
\title[Wavelet analysis of matter clustering]{Continuous wavelet analysis of matter clustering using the Gaussian-derived wavelet}
\author[0000-0003-4064-417X]{Yun Wang}
\affiliation{College of Physics, Jilin University, Changchun 130012, P.R. China.}
\author[0000-0002-0586-7178]{Hua-Yu Yang}
\affiliation{College of Physics, Jilin University, Changchun 130012, P.R. China.}
\author[0000-0001-7767-6154]{Ping He}
\affiliation{College of Physics, Jilin University, Changchun 130012, P.R. China.}
\affiliation{Center for High Energy Physics, Peking University, Beijing 100871, P.R. China.}
\correspondingauthor{Ping He} \email{hep@jlu.edu.cn}
% ------------------------------------------------------------------------------------------------
\begin{abstract}
Continuous wavelet analysis has been increasingly employed in various fields of science and engineering due to its remarkable ability to maintain optimal resolution in both space and scale. Here, we introduce wavelet-based statistics, including the wavelet power spectrum, wavelet cross-correlation and wavelet bicoherence, to analyze the large-scale clustering of matter. For this purpose, we perform wavelet transforms on the density distribution obtained from the one-dimensional Zel'dovich approximation and then measure the wavelet power spectra and wavelet bicoherences of this density distribution. Our results suggest that the wavelet power spectrum and wavelet bicoherence can identify the effects of local environments on the clustering at different scales. Moreover, we apply the statistics based on the three-dimensional isotropic wavelet to the IllustrisTNG simulation at $z=0$, and investigate the environmental dependence of the matter clustering. We find that the clustering strength of the total matter increases with increasing local density except on the largest scales. Besides, we notice that the gas traces the dark matter better than stars on large scales in all environments. On small scales, the cross-correlation between the dark matter and gas first decreases and then increases with increasing density. This is related to the impacts of the AGN feedback on the matter distribution, which also varies with the density environment in a similar trend to the cross-correlation between the dark matter and gas. Our findings are qualitatively consistent with previous studies about the matter clustering.
\end{abstract}

%----------------------------------------------------------------------------------------------

% ----------------------------------------------------------------------------------------------
\section{Introduction}
\label{sec:intro}

The standard cosmological model ($\Lambda$CDM) states that the hierarchical matter clustering evolved from the random primordial fluctuations via gravitational instability, as revealed by the observed pattern of spatial distribution for luminous objects \citep{Tegmark2004}. However, luminous objects are made up of baryons, whose clustering is not completely consistent with that of dark matter (DM) component because of a series of physical processes such as radiative cooling, star formation and feedback processes. With hybrid $N$-body/hydrodynamic simulations, the baryonic effects on the large-scale clustering of matter have been extensively studied in Fourier space, and these studies show that the deviation between baryons and dark matter in mass distribution depends on the scale and redshift \citep[e.g.][]{Chisari2019, vanDaalen2020, Yang2020}. Moreover, many studies indicate that the clustering of both dark matter and baryonic matter is environment-dependent \citep[e.g.][]{Abbas2005, Peng2010, Wang2018, Man2019}. However, usual statistical schemes based on the ordinary Fourier transform, e.g. power spectrum and bispectrum, cannot be used to measure the environmental and scale dependence of the matter clustering at the same time, since these schemes do not contain any positional information of the signal. The straightforward solution to overcome this problem is to use techniques of the windowed Fourier transform (WFT), which takes harmonic waves multiplied by a window of fixed width as basis functions, hence enabling us to gain the positional and scale information simultaneously \citep{Gabor1946}.

However, the fixed window size of the WFT leads to a bad space-scale resolution \citep[e.g.][]{Kaiser1994,Romeo2004,Gao2011}. If we set the window size to be small enough, then the spatial resolution certainly matches the signal, but the scale/frequency localization is too poor to resolve the large scales/low frequencies. If we choose a wider window in order to have a finer scale/frequency resolution, the spatial resolution becomes too coarse to detect small scale structures. Therefore, the WFT is not a suitable tool for analyzing signals with a wide range of scales, such as spatial distributions of the dark matter and galaxies.

An alternative way to achieve simultaneous analysis of space and scale is the wavelet transform, which provides an adaptive space-scale resolution \citep[e.g.][]{Daubechies1992,Kaiser1994,Chui1997,Torrence1998,Van den Berg2004,Mallat2009,Addison2017}. Wavelet transform analysis decomposes a signal into separate scale components using a set of scaled and shifted wavelets, which are well localized in both real and Fourier space. Consequently, the local features of the signal at different scales are revealed by this decomposition. Wavelet transform methods can be classified into the discrete wavelet transform (DWT) and the continuous wavelet transform (CWT). The DWT, using orthogonal wavelet bases, operates over scales and positions based on the integer power of two, hence giving the most compact representation of the signal. This leads to the effectiveness and ease of implementation of the DWT, which hence is particularly useful for information compression \citep[e.g.][]{Khaliffa2008, Abdulazeez2020}. Due to its advantages, the DWT has also been applied to study the large-scale structure of the universe \citep[e.g.][]{Fujiwara1996,Pando1996,Pando1998, Pando2004, Fang2000,Romeo2004, Liu2008, Lu2010}. However, there are mainly two drawbacks in the DWT caused by dyadic scales. Firstly, the DWT provides the poor scale resolution so that some meaningful features of a signal cannot be detected. Secondly, it lacks translational invariance. The so-called translational invariance means that if a signal is translated, then its wavelet coefficients is translated by the same amount without other modification at every scales \citep{Addison2017}. Obviously it is not the case for the DWT. A small translation on the signal can make the discrete wavelet coefficients vary substantially on different scales, thereby the total energy in the wavelet domain being not conserved after the signal was shifted. These drawbacks suggest that the DWT is not suitable for analyzing signals with extremely high complexity.

In contrast to the DWT, the CWT allows the scale and translation parameters to continuously change, which makes it translational invariant and redundant. The redundancy guarantees that the CWT can provide high resolution results, which are much easier to interpret than those obtained with the DWT \citep{Aguiar2014, Addison2018}. As a result, the CWT is becoming more and more popular across different disciplines, including geophysics, biomedicine, economics, astrophysics, fluid mechanics and so on. For instance in astrophysics and cosmology, the CWT is used for the detection of 21cm signal \citep[e.g.][]{Gu2013}, the detection of baryonic acoustic oscillations \citep[e.g.][]{Tian2011, Arnalte-Mur2012, Labatie2012}, the detection of substructures in two-dimensional (2D) mass maps \citep[e.g.][]{Flin2006, Schwinn2018}, the analysis of turbulence evolution in the intracluster medium \citep[e.g.][]{Shi2018, Roh2019}, the correlation analysis of galactic images \citep[e.g.][]{Frick2001, Tabatabaei2013}, the analysis of the multifractal character of the galaxy distribution \citep[e.g.][]{Martinez1993, Rozgacheva2012} and the analysis of the CMB maps \citep[e.g.][]{Cayon2001, Starck2004, Gonzalez2006, Curto2011}.

The main problem faced by the CWT, e.g. when analyzing one-dimensional (1D) signals, is that its classical inverse formula is a double integration, which results in a heavy computational effort for recovering the original signal. Although there is an alternative inverse transform formula in the form of single-integral for the complex-valued wavelets, the real wavelets has generally been considered to have no such simple inverse transformation \citep{Delprat1992, Aguiar2014}. In our previous work \citep{Wang2021}, we proposed a novel scheme of constructing continuous wavelets to overcome this problem, in which the wavelet functions are obtained by taking the first derivative of smoothing functions with regard to the positive defined scale parameter. With this scheme, the original signal is recovered easily by integrating the wavelet coefficients with respect to the scale parameter. As an inspired example, we took the Gaussian function as a smoothing function to derive the wavelet dubbed the {\em Gaussian-derived wavelet} (hereafter GDW) and briefly discussed its preliminary application to the matter power spectrum, demonstrating the success of this scheme.

In this work, we use local statistics established on wavelet coefficients to further investigate the potential of GDW for analyzing the large-scale clustering of matter. Specifically, the wavelet power spectrum (WPS), the wavelet cross-correlation (WCC), and the wavelet bicoherence (WBC) are employed here. The WPS and WCC were first introduced by \citet{Hudgins1993} to examine atmospheric turbulence. The WPS measures the variance of a signal at various scales within some local region, and the WCC is used to quantify similarities between two signals. The WBC was originally proposed by \citet{Milligen1995a, Milligen1995b} to detect the short-lived structures induced by phase coupling in turbulence. Moreover, the cosmic baryonic density distribution at late times is similar to a fully developed turbulence \citep{Shandarin1989,He2006}, which enlightens us to apply these tools to the context of structure formation of the universe. To illustrate our approach in a more intuitive way, we shall use the 1D matter distributions, since a great deal of work on the large-scale structure of the universe is accomplished using 1D cosmology models \citep{Gouda1989, Fujiwara1996, Tatekawa2001, Miller2010, Manfredi2016}. The Zel'dovich approximation is a simple model that provides a good approximated solution for the nonlinear evolution of collisionless matter \citep{Zeldovich1970}. In the 1D case, the Zel'dovich approximation is proved to be an exact solution in the fully nonlinear regime until the first singularity appears \citep{Soda1992}. It is straightforward and efficient to calculate the 1D Zel'dovich solution, and so we discuss our analysis method with it. Firstly, we decompose the density fields obtained from the 1D Zel'dovich approximation into wavelet components at different positions and scales. Then, by measuring the WPS and WBC of matter densities based on these components, we investigate the effects of density environment on the matter clustering. 

To further demonstrate the ability of wavelet statistics in characterizing the environmental dependence of the matter clustering, we generalize them to the three-dimensional (3D) versions and then apply them to the 3D density fields for different matter components in the IllustrisTNG simulation \citep{Springel2018, Naiman2018, Nelson2018, Pillepich2018, Marinacci2018}. By measuring the WPS and WCC, we can investigate (1) the variation of the clustering strength with the density environment at given scales, (2) the variation of the cross-correlations between different matter components with the density environment at given scales and (3) the baryonic effects on the matter clustering in different density environments.

This paper is organized as follows. In section 2, we describe the characteristic of the GDW, the theory of the CWT and wavelet-based statistical tools. In section 3, we briefly introduce the data we used, including 1D Zel'dovich approximation and density fields in IllustrisTNG. In section 4, we give the numerical results for the matter clustering. Finally in section 5, we discuss and summarize our results.

%---------------------------------------------------------------------------------
\section{Methods of continuous wavelet analysis}
\label{sec:methods_cw}

In this section, we briefly review the definition of the CWT based on the GDW in 1D and then introduce the wavelet-based statistical tools, i.e. the WPS, the WCC function and the WBC. To assess their significance, statistical errors are then discussed. Finally, we give the correspondence between the wavelet scale and the Fourier wavenumber for the GDW. The 3D isotropic CWT and the statistics formulated on it are given in Appendix \ref{sec:iso_CWT}.

%------------------------------------------------------------------------
\begin{table}
	\centering
	\caption{Notations used in the paper, with their meanings and Acronyms. In this table, we only list the symbols associated with the 1D CWT. The same notation convention is also used for the quantities based on the 3D isotropic CWT, except that the scalars $x$ and $k$ are replaced with vectors $\mathbf{r}$ and $\mathbf{k}$, and the length $L$ is replaced with volume $V$. }
	\label{tab:notations}
	\begin{tabular}{lll}
		\toprule
		Notation & Meaning & Acronym  \\ \\[-1.1em]
		\hline
		$\psi(w,x)$          & Gaussian-derived wavelet        & GDW     \\ \\[-0.5em]
		$\hat\psi(w,k)$      & Fourier transform of the GDW    &         \\ \\[-0.5em]
		$W_f(w,x)$           & wavelet transform               & WT      \\ \\[-0.5em]
		$\hat W_f(w,k)$      & Fourier transform of the WT     &         \\ \\[-0.5em]
		$P_{f,L}^{W}(w)$     & wavelet power spectrum          & WPS     \\ \\[-0.5em]
		$P^F_f(k)$           & Fourier power spectrum          &         \\ \\[-0.5em]
		$XWT_{f-g}(w,x)$        & cross-wavelet transform      & XWT     \\ \\[-0.5em]
		$C_{f-g,L}^{W}(w)$      & wavelet cross-correlation    & WCC     \\ \\[-0.5em]
		$B_{f,L}^{W}(w_1, w_2)$ & wavelet bispectrum           & WBS     \\ \\[-0.5em]
		$b_{f,L}^{W}(w_1, w_2)$ & wavelet	bicoherence        & WBC     \\ \\[-0.5em]
		$b_{f,L}^{W}(w)$        & summed WBC                   &         \\ \\[-0.5em]
		$b_{f,L}^{W}$           & total WBC                    &         \\ \\[-1.0em]
		\hline
	\end{tabular}
\end{table}

%---------------------------------------------------------------------------------------------------------------------
\subsection{The Gaussian-derived wavelet and the continuous wavelet transform}
\label{sec:gdw}

\begin{figure}
	\centerline{\includegraphics[width=0.45\textwidth]{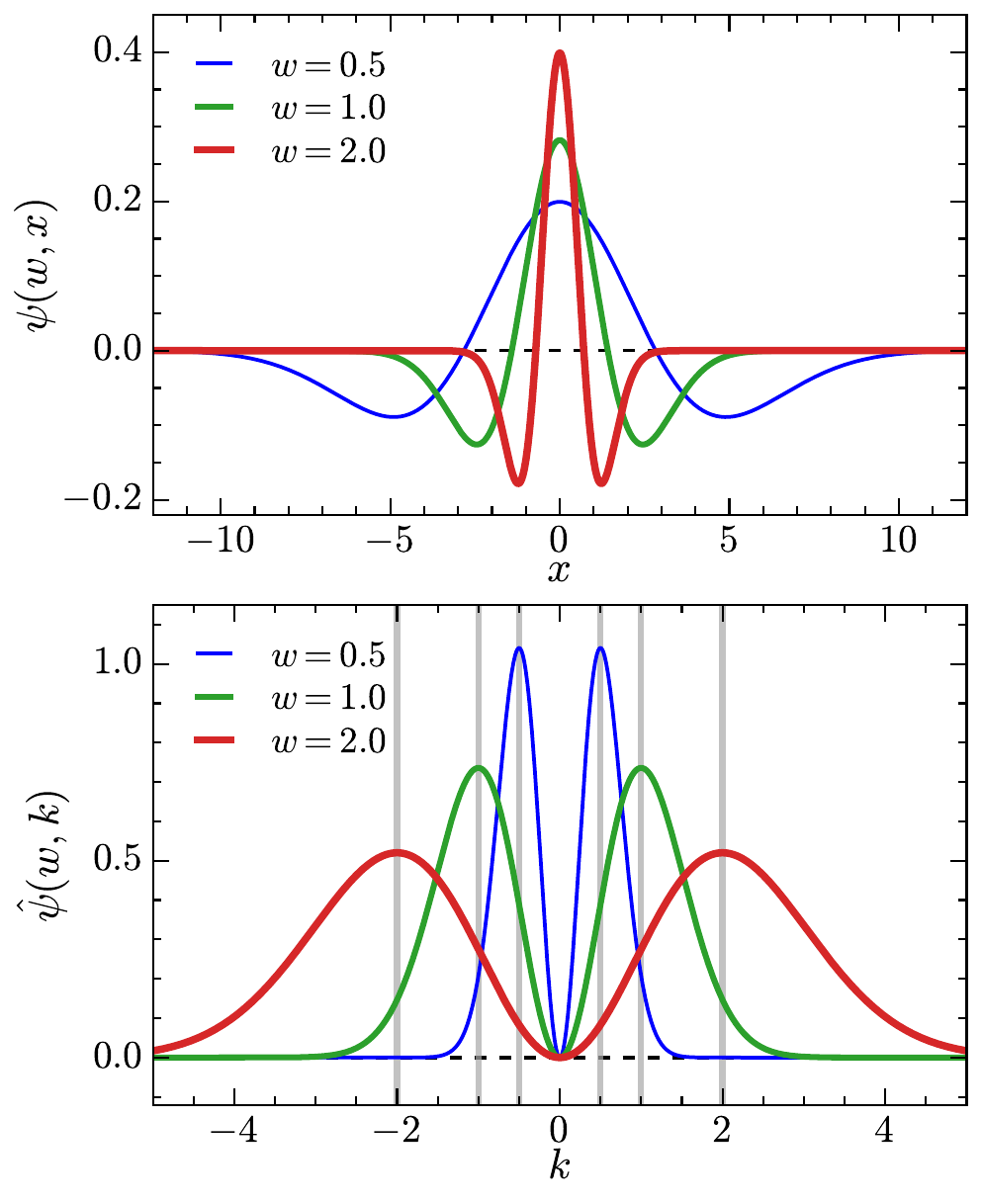}}
	\caption{The Gaussian-derived wavelet, $\psi(w,x)$ (top) and the corresponding Fourier transform $\hat\psi(w,k)$ (bottom) for three scale parameters, $w=0.5, 1.0$ and $2.0$. The grey vertical lines denote wavenumbers where $\hat\psi(w,k)$ takes the maximum.}
	\label{fig:GaussianWavelet}
\end{figure}

The GDW, used in this work, is defined as the first derivative of the Gaussian smoothing function with respect to the scale parameter $w$, multiplied by a factor $\sqrt{w}$,
% --------------------------------------------------------------------------------------------------------------------
\begin{flalign}
\label{eq:GaussianWavelet}
\psi(w,x) &\equiv \sqrt{w}\frac{\partial G(w,x)}{\partial w} \nonumber\\
& = \frac{\sqrt{w}}{4\sqrt{\pi}}(2-w^2x^2)\exp\left(-\frac{w^2x^2}{4}\right), &&
\end{flalign}
where $G(w,x)=w\exp\left(-w^2x^2/4 \right)/(2\sqrt{\pi})$ denotes the Gaussian smoothing function with the scale parameter $w$ greater than zero. The prefactor $\sqrt{w}$ on the right hand side of equation~(\ref{eq:GaussianWavelet}) ensures that the energy of the wavelet is unaffected by the scale parameter.\footnote{In the previous work \citep{Wang2021}, we simply define GDW as $\psi(w,x) \equiv \partial G(w,x)/\partial w$.} The Fourier transform of the GDW is given by
% ---------------------------------------------------------------------------------------------------------------------
\begin{flalign}
\label{eq:FTofGW}
\hat\psi(w,k)=\frac{2}{\sqrt{w}}\left(\frac{k}{w}\right)^2\exp\left[-\left(\frac{k}{w}\right)^2 \right], &&
\end{flalign}
through which the scale parameter $w$ is defined as the peak wavenumber of $\hat\psi(w,k)$, i.e. the wavenumber at which $\hat\psi(w,k)$ takes the maximum value, as demonstrated in Fig.~\ref{fig:GaussianWavelet}. It can be seen from equation~(\ref{eq:GaussianWavelet}) and (\ref{eq:FTofGW}) that the GDW satisfies the conditions of admissibility, similarity and regularity. In fact, the GDW is the same as the Mexican hat wavelet in 1D, except for their constant factors. This is illustrated by substituting $w=\sqrt{2}/a$ into equation~(\ref{eq:GaussianWavelet}), where $a$ is the scale parameter for the traditional CWT.\footnote{The detailed comparison between the GDW and the Mexican hat wavelet is given in \citet{Wang2021}.} However, as pointed out by \citet{Wang2021}, the 3D GDW is an anisotropic separable wavelet function, which is not a 3D Mexican hat wavelet. At present, we concern only the 1D case.

The wavelet function $\psi(w,x)$ is then used to perform the wavelet transform of a 1D signal $f(x)$ as follows
% --------------------------------------------------------------------------------------------------------------------
\begin{flalign}
\label{eq:CWT}
W_f(w,x) = \int_{-\infty}^{+\infty}f(u)\psi(w,x-u)\mathrm{d}u, &&
\end{flalign}
Note that there is no complex conjugate in equation~(\ref{eq:CWT}) since both signal and wavelet are real. The wavelet transform, equation~(\ref{eq:CWT}), is nothing but a convolution of the signal with wavelets at different scales, which can be implemented efficiently by FFT technique. As a function of scale $w$ and space $x$, from the wavelet transform we can clearly see how different scale features are localized in space. However, it is not possible to achieve arbitrarily good resolution in space and scale simultaneously \citep{Chui1997, Addison2017}. As illustrated in Fig.~\ref{fig:GaussianWavelet}, a narrower (wider) wavelet provides better (poorer) spatial resolution accompanied by poorer (better) frequency resolution. This fact is quantified in terms of the uncertainty principle $\Delta x\Delta k\gtrsim1/2$ for the GDW, where $\Delta x$ is the standard deviation of the wavelet in real space and $\Delta k$ is the standard deviation in frequency \citep{Chui1997}.

The greatest convenience brought to us by the definition of GDW is that the signal can be reconstructed through a single integral of the wavelet coefficients. Combining equation~(\ref{eq:GaussianWavelet}) and (\ref{eq:CWT}), we have
% ---------------------------------------------------------------------------
\begin{flalign}
\label{eq:wt&smoothedfield}
W_f(w,x)=\sqrt{w}\frac{\partial f_{\rm s}(w,x)}{\partial w}, &&
\end{flalign}
where $f_{\rm s}(w,x)$ refers to the smoothed field under the scale $w$ and is given by
% ---------------------------------------------------------------------------
\begin{flalign}
\label{eq:smoothedfield}
f_{\rm s}(w,x)=\int_{-\infty}^{+\infty}f(u)G(w,x-u)\mathrm{d}u.&&
\end{flalign}
By integrating equation~(\ref{eq:wt&smoothedfield}) with respect to $w$, we get the single integral inverse transform as shown below,
% ---------------------------------------------------------------------------
\begin{flalign}
\label{eq:inverse_cwt}
f(x) = C + \int_0^{+\infty} \frac{W_f(w,x)}{\sqrt{w}}\mathrm{d}w,&&
\end{flalign}
where the integral constant $C$ is equal to $f_{\rm s}(w=0,x)$, and generally $C=0$ for, say, the density contrast field of the Universe. Compared with the usual reconstruction formula for the CWT found in most wavelet literatures, e.g. \citet{Addison2017}, our reconstruction formula defined by equation~(\ref{eq:inverse_cwt}) is much easier to compute numerically and generalize to integrations of higher dimensions.

%-------------------------------------------------------------------------------

% -----------------------------------------------------------------------------
\subsection{Parseval's theorem for the continuous wavelet transform}
\label{sec:Parseval}

Parseval's theorem is an important result in Fourier Transform, which states that the inner product between signals is preserved in going from time to the frequency domain. Similarly, there is an analogue of Parseval's theorem for the wavelet transform \citep{Hudgins1993}, the form of which is expressed as
% ---------------------------------------------------------------------------
\begin{flalign}
\label{eq:wavelet_parseval_a}
&\int_0^{+\infty}\int_{-\infty}^{+\infty}W_f(w,x)W_g(w,x)\mathrm{d}x\mathrm{d}w  \nonumber\\
& = C_\psi\int_{-\infty}^{+\infty}f(u)g(u)\mathrm{d}u, &&
\end{flalign}
where $W_f(w,x)$ and $W_g(w,x)$ are WTs for $f(x)$ and $g(x)$ respectively, $C_\psi = \int_0^{+\infty} (\hat\psi(1,k)^2/k)\mathrm{d}k$ is the admissibility constant, and equals to $1/2$ for the GDW. If $f$ and $g$ are the same, then we have
% ---------------------------------------------------------------------------
\begin{flalign}
\label{eq:wavelet_parseval_b}
\int_0^{+\infty}\int_{-\infty}^{+\infty}|W_f(w,x)|^2\mathrm{d}x\mathrm{d}w = C_\psi \int_{-\infty}^{+\infty}|f(u)|^2\mathrm{d}u.&&
\end{flalign}

%------------------------------------------------------------------------------------------------
\subsection{The wavelet power spectrum and the wavelet cross-correlation}
\label{sec:wps&wc}

We assume that the signal $f(x)$ satisfies the periodic boundary condition with period $L_\mathrm{b}$, which is a usual choice for studying a typical region of the universe. Then for a sub-region of the signal $L\leqslant L_\mathrm{b}$, the local WPS is defined as
% -----------------------------------------------------------------------------------------------
\begin{flalign}
\label{eq:wps}
P_{f,L}^{W}(w)=\frac{1}{L}\int_L |W_f(w,x)|^2\mathrm{d}x. &&
\end{flalign}
We can see that equation~(\ref{eq:wps}) refers to the variance at the scale $w$ within the spatial region $L$, since $|W_f(w,x)|^2$ just the variance per area at the space-scale plane, as referred to by the Parseval's theorem, equation~(\ref{eq:wavelet_parseval_b}). According to the Fourier Parseval's theorem for periodical signals, we can obtain the relationship between the global wavelet and Fourier power spectrum, which is given by
% ---------------------------------------------------------------------------
\begin{flalign}
\label{eq:wps&fps}
P_{f,L_b}^{W}(w) &=\frac{1}{L_b}\int_0^{L_b}|W_f(w,x)|^2\mathrm{d}x \nonumber\\
& = \frac{1}{L_b}\sum_k \frac{|\hat W_f(w,k)|^2}{L_b} \nonumber \\
& = \sum_k P^F_\psi(w,k)P^F_f(k),&&
\end{flalign}
where $\hat W_f(w,k)=\hat \psi(w,k)\hat f(k)$ is the Fourier transform of $W_f(w,x)$, $P^F_f(k)=|\hat f(k)|^2/L_\mathrm{b}$ is the Fourier power spectrum of the signal, and $P^F_\psi(w,k)= |\hat\psi(w,k)|^2/L_\mathrm{b}$ denotes the Fourier power spectrum of the wavelet. Obviously, the global WPS of a signal is the average of its Fourier power spectrum weighted by the Fourier power spectrum of the corresponding wavelet function over all wavenumbers.

Then given two signals $f(x)$ and $g(x)$ with wavelet transforms $W_f(w,x)$ and $W_g(w,x)$, we can define the XWT as
% ---------------------------------------------------------------------------
\begin{flalign}
\label{eq:xwt}
\mathit{XWT}_{f-g}(w,x) = W_f(w,x)W_g(w,x),&&
\end{flalign}
i.e. it measures the local covariance at each spatial position and scale, as revealed by equation~(\ref{eq:wavelet_parseval_a}). By integrating the XWT over a finite spatial region, we get the normalized local WCC as follows
% ---------------------------------------------------------------------------
\begin{flalign}
\label{eq:wc}
C^{W}_{f-g,L}(w)=\frac{1}{L}\frac{\int_L \mathit{XWT}_{f-g}(w,x)\mathrm{d}x}{(P_{f,L}^{W}P_{g,L}^{W})^{1/2}},&&
\end{flalign}
which can take on values between $-1$ (perfect anti-correlation) and $1$ (perfect correlation).

%--------------------------------------------------------------------------------------------------
\subsection{The wavelet bicoherence and error estimation}
\label{sec:wb}

The main content of this sub-section is based on \citet{Milligen1995a, Milligen1995b}, and we refer the interested readers to these two references for detail.

The Fourier bispectrum is the lowest order statistic that measures the amount of phase-coupling of harmonic modes within a signal. By analogy to it, the WBS is given as
%--------------------------------------------------------------------------------------------------
\begin{flalign}
\label{eq:wbs}
B^{W}_{f,L}(w_1,w_2)= \frac{1}{L}\int_LW_f(w_1,x)W_f(w_2,x)W_f(w,x)\mathrm{d}x,&&
\end{flalign}
where $w_1+w_2=w$ (frequency sum rule). The WBS measures the non-linear interplay within the local region $L$ between scale components $w_1$, $w_2$ and $w$ such that the sum rule is satisfied. In the case of completely random phases of the signal, $B^{W}_{f,L}(w_1,w_2)$ is statistically to be zero. However, once a coherent structure is formed by the phase-coupling, $B^{W}_{f,L}(w_1,w_2)$ will take significant non-zero values. The WBS usually is normalized in the following way:
% ---------------------------------------------------------------------------
\begin{flalign}
\label{eq:wbc}
b^{W}_{f,L} & (w_1,w_2) = \nonumber \\ 
&\left[\frac{|B^{W}_{f,L}(w_1,w_2)|^2}{\frac{1}{L}\int_L|W_f(w_1,x)W_f(w_2,x)|^2\mathrm{d}x \ P^{W}_{f,L}}\right]^{1/2},&&
\end{flalign}
which is called the WBC, attaining values between 0 and 1. Throughout this paper, we will use it to measure the non-linear behaviors of matter clustering instead of the WBS. In addition, it is convenient to introduce the summed WBC defined as
%----------------------------------------------------------------------------
\begin{flalign}
\label{eq:summed_wbc}
b^{W}_{f,L}(w)=\left[ \frac{1}{s(w)}\sum_{w_1,w_2} [b^{W}_{f,L}(w_1,w_2)]^2  \right]^{1/2}, &&
\end{flalign}
where the summation is taken over all $w_1$ and $w_2$ such that $w_1+w_2$ equals to $w$ and $s(w)$ is the number of summands in the summation. In addition, as a measure of the total non-linearity in the chosen region of the signal, the total WBC is defined by averaging the squared WBC over all points in the scale-scale plane as
%----------------------------------------------------------------------------
\begin{flalign}
\label{eq:total_wbc}
b^{W}_{f,L} = \left[ \frac{1}{S} \sum_{w_1,w_2}[b^{W}_{f,L}(w_1,w_2)]^2  \right]^{1/2}, &&
\end{flalign}
where $S$ is the total number of points $(w_1,w_2)$ in the scale-scale plane.

In practice for discrete sampled signals, integrations over the interval $L$ involved in calculating the statistics mentioned above are carried out with summation over $N$ sample points. By the law of large numbers, an integration over $L$ suffers a relative statistical error of $1/\sqrt{N}$. In addition, the fact that CWTs are non-orthogonal leads to that wavelet coefficients are not all statistically independent \citep{Milligen1995a, Milligen1995b, Milligen1997}. If continuous wavelets $\psi(x)$ and $\psi^*(x+d_0)$ satisfy $\int \psi(x)\psi^*(x+d_0)\mathrm{d}x=0$, then they can be regarded as approximately orthogonal and corresponding wavelet coefficients are statistically independent. In the case of GDW, independent wavelet coefficients are separated by distance $d_0/w$ at each scale, where $d_0 = 2 \sqrt{3-\sqrt{6}}$. Then the number of statistical independent points on the interval $L$ is $N'=L/(d_0/w)=2\pi N w/(d_0 k_\mathrm{samp})$, where $k_\mathrm{samp} = 2\pi/\Delta x$ is the sampling frequency. Thus for the WPS, its statistical error is estimated by
%-------------------------------------------------------------------------------
\begin{flalign}
\label{eq:wp_error}
\epsilon\left[P_{f,L}^{W}(w)\right] \approx \frac{P_{f,L}^{W}(w)}{\sqrt{N'}} = P_{f,L}^{W}(w) \left(\frac{ d_0 k_\mathrm{samp}}{2\pi N w}\right)^{1/2}. &&
\end{flalign}
By applying similar estimates for all integral terms in equation~(\ref{eq:wc}) and according to error propagation, we obtain the statistical noise level for the WCC,
%-------------------------------------------------------------------------------
\begin{flalign}
\label{eq:wc_noise}
\epsilon\left[|C^{W}_{f-g,L}(w)|\right] \approx  \left( \frac{3 d_0 k_\mathrm{samp}}{4\pi N w} \right)^{1/2}. &&
\end{flalign}
Equation~(\ref{eq:wc_noise}) is called the statistical noise level, because it is the cross-correlation value that can be achieved by a Gaussian noise and is caused by using a limited number of values in the integration. Notice that this noise level is scale-dependent, suggesting that the interpretation of the signal becomes increasingly significant as the scale decreases. Just like the approach above, we can obtain the noise level of the WBC shown below,
%-----------------------------------------------------------------------------
\begin{flalign}
\label{eq:wbc_noise}
\epsilon\left[b^{W}_{f,L}(w_1,w_2)\right]\approx \left( \frac{d_0 k_\mathrm{samp}}{2\pi N \mathrm{Min}(w_1,w_2,w_1+w_2)} \right)^{1/2}.&&
\end{flalign}

%------------------------------------------------------------------------------
\subsection{The relationship between scale and wavenumber}
\label{sec:scale&wavenum}

\begin{figure}
	\centerline{\includegraphics[width=0.5\textwidth]{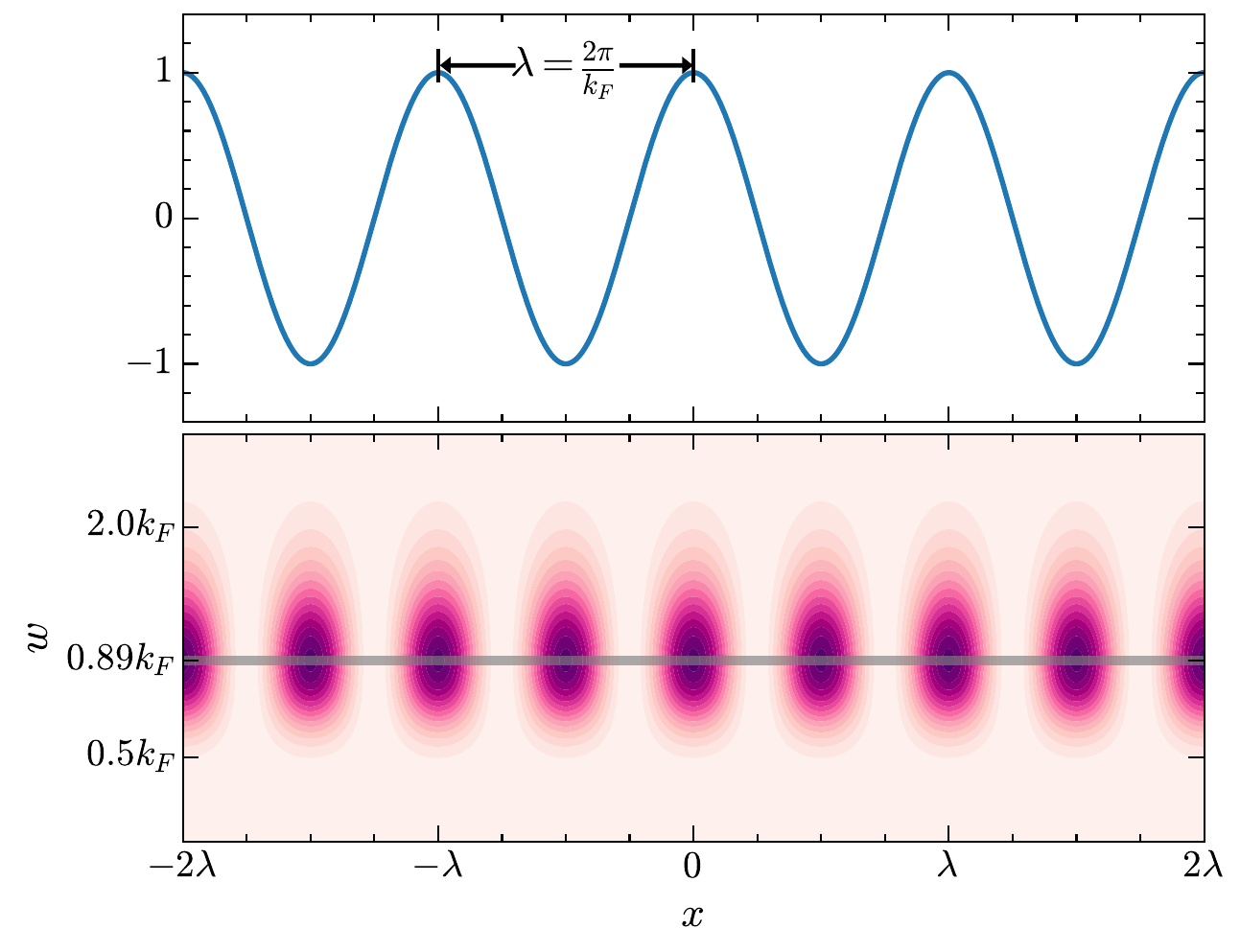}}
	\caption{The squared wavelet transform plot of a cosine function. Upper panel: a cosine wave of period $\lambda=2\pi/k_F$. Lower panel: the corresponding contour plot of $|W_\mathrm{cos}(w,x)|^2$ for the cosine wave. The gray horizontal line indicates that $|W_\mathrm{cos}(w,x)|^2$ takes its maximum value at $w \approx 0.89k_F$.}
	\label{fig:scale&wavenum}
\end{figure}

To facilitate the comparison between wavelet and Fourier spectra, we need to ascertain the relationship between the scale parameter $w$ and the equivalent Fourier wavenumber. \citet{Meyers1993} and \citet{Torrence1998} suggest that the relationship between them can be derived analytically for a particular wavelet function by performing wavelet transform of a cosine wave with known period, such as $\cos{(k_Fx)}$, and then computing the scale $w$ at which the scalogram reaches its maximum. Following their method, the scalogram of $\cos{(k_Fx)}$ is first computed with the GDW, and the result is
% ---------------------------------------------------------------------------
\begin{flalign}
\label{eq:scalo_cos}
|W_\mathrm{cos}(w,x)|^2=\frac{4 k_F^4 \cos^2(k_F x) \exp{(-\frac{2 k_F^2}{w^2})}}{w^5}, &&
\end{flalign}
which is depicted in Fig.~\ref{fig:scale&wavenum}. The scale parameter that makes $|W_\mathrm{cos}(w,x)|^2$ take the maximum value should be equivalent to the wavenumber $k_F$, since the Fourier power spectrum of a cosine wave is an impulse at $k_F$. Therefore, by solving $\partial |W_\mathrm{cos}(w,x)|^2/\partial w=0$, we get the correspondence between the wavelet scale and Fourier wavenumber,
% ---------------------------------------------------------------------------
\begin{flalign}
\label{eq:scale&wavenum}
w =\frac{2}{\sqrt{5}}k_F\approx 0.89k_F.&&
\end{flalign}
The relation of equation~(\ref{eq:scale&wavenum}) shows that the wavelet scale is proportional to the Fourier wavenumber.\footnote{Following the same logic, the relation between the wavelet scale and its equivalent Fourier wavenumber is $w=2k/\sqrt{7}$ for the 3D isotropic CWT.} In the next sections, we will present our results in terms of the equivalent Fourier wavenumber $k_F$ instead of $w$, and we drop the subscript `$F$' for simplicity of notation.

% ---------------------------------------------------------------------------

For convenience of readers, in Table~\ref{tab:notations}, we list all the notations used in our paper, with their meanings and the corresponding acronyms.

%---------------------------------------------------------------------------------
\section{Data sets}
\label{sec:data}

In this work, we use two types of data, i.e. the 1D density fields obtained from Zel'dovich approximation and 3D density fields of the IllustrisTNG simulations. In the 1D case, Zel'dovich approximation provides the exact non-linear solution for the perturbative equations of collisionless matter up to the first appearance of orbit-crossing singularities \citep{Soda1992}. Due to easy implementation and fast computation of the 1D Zel'dovich exact solution, we will demonstrate the usefulness of our method with it before further analyzing the cosmological simulation IllustrisTNG.

% ---------------------------------------------------------------------------
\begin{figure}
	\centerline{\includegraphics[width=0.5\textwidth]{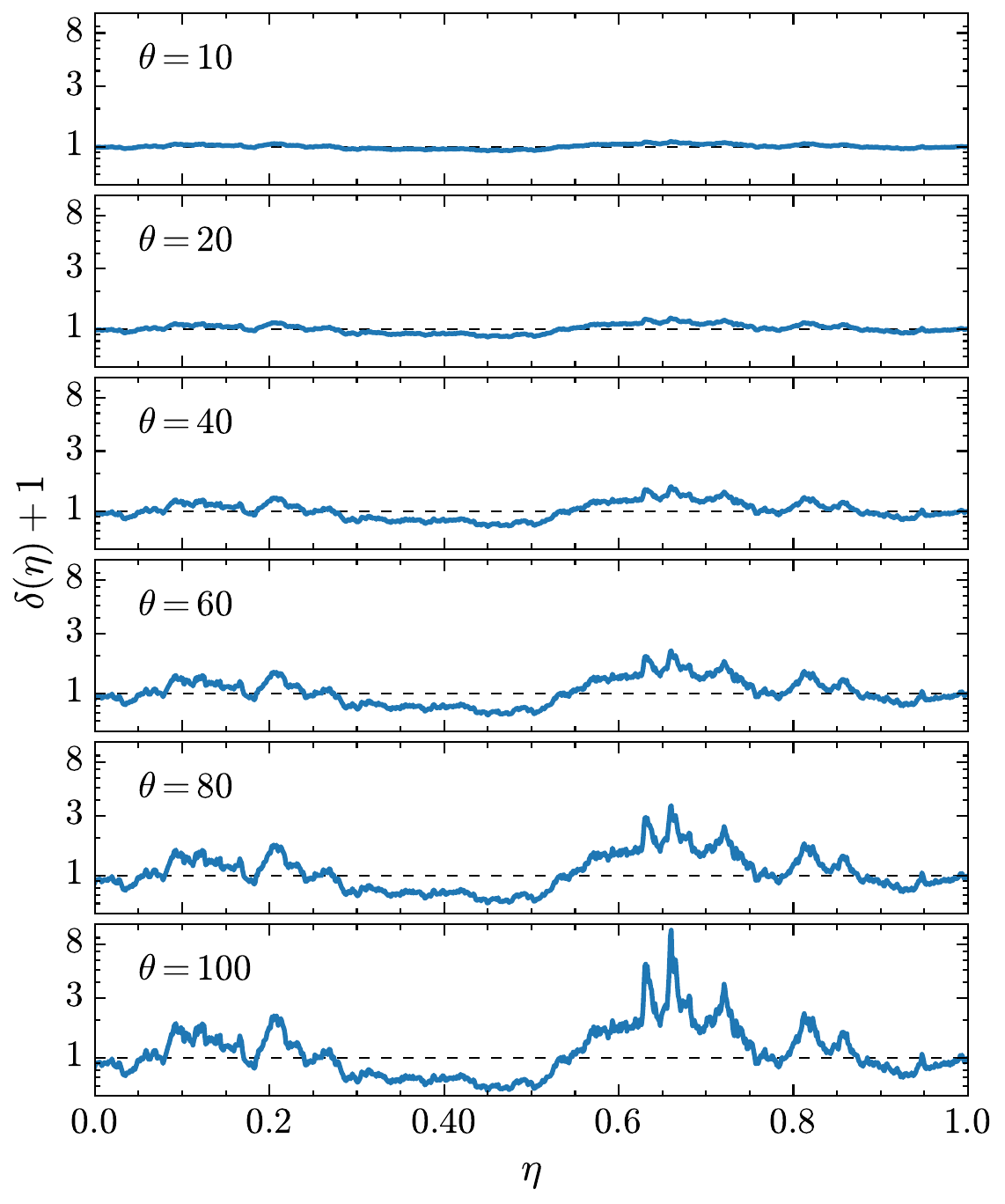}}
	\caption{The 1D density fields obtained from Zel'dovich approximation at times $\theta=10, 20, 40, 60, 80$ and $100$ from top to bottom.}
	\label{fig:zel_dens}
\end{figure}

%---------------------------------------------------------------------------------
\subsection{Zel'dovich approximation in one dimension}
\label{sec:zel_approx}

The fundamental idea of Zel'dovich approximation is the transformation between Eulerian and Lagrangian coordinates, i.e.
%----------------------------------------------------------------------------
\begin{flalign}
\label{eq:eluer&lagran}
x(t,q)=q-\theta(t)f(q), &&
\end{flalign}
where $x$ and $q$ are the Eulerian and Lagrangian coordinates, respectively. Then by applying the mass conservation to equation~(\ref{eq:eluer&lagran}), the density contrast is given explicitly by
%---------------------------------------------------------------------------------------------------------------------
\begin{flalign}
\label{eq:zel_dens}
\delta \left( \eta,t \right)+1 =\frac{1}{1-\theta\left( t \right) F\left( \eta \right)}, &&
\end{flalign}
where $F(\eta)=\frac{\mathrm{d}f(\eta L_b)}{\mathrm{d}\eta}/L_b$ with $\eta=q/L_b$ being the dimensionless coordinates divided by the length size of density field, and $\theta(t)$ is the growth factor. For simplicity, we normalize $\theta(t)$ to unity at the initial time and use it as a time variable instead of $t$. It is easy to see from equation~(\ref{eq:zel_dens}) that $F(\eta)$ is the initial density contrast if small enough. For more details about the 1D Zel'dovich approximation, we refer the readers to \citet{Soda1992} and \citet{Fujiwara1996}.

The initial condition is set to be
% ------------------------------------------------------------------------------------------------------------------
\begin{flalign}
\label{eq:zel_ic}
F(\eta)=2\sum_{k>0}\sqrt{P_\mathrm{i}(k)}[B_k\cos(k\eta)+C_k\sin(k\eta)],&&
\end{flalign}
where $B_k$ and $C_k$ are drawn from Gaussian with standard deviation of 1. We impose periodic boundary condition on the interval $\eta\in[0,1]$, then divide it into 1024 equally spaced segments. So the wave number, as an integer multiple of $2\pi$, has a maximum value of $512\cdot2\pi$, i.e. the Nyquist frequency. In this work, we assume that the spectral index of the initial power spectrum $P_\mathrm{i}(k)=Ak^{n}$ in equation~(\ref{eq:zel_ic}) is equal to $n = -2$.  The amplitude $A$ is chosen to be $2.5\times10^{-6}$ such that the initial density perturbation is between $-0.01$ and $0.01$. Hence evolution of the density field is totally determined by equation~(\ref{eq:zel_dens}) and (\ref{eq:zel_ic}).

As shown in Fig.~\ref{fig:zel_dens}, we select the density fields at $\theta = 10$, $20$, $40$, $60$, $80$, and $100$ to examine their evolution from linear to non-linear stages.

%------------------------------------------------------------------------------
\subsection{IllustrisTNG data}

The IllustrisTNG project is a suite of state-of-the-art cosmological hydrodynamic simulations \citep{Springel2018, Naiman2018, Nelson2018, Pillepich2018, Marinacci2018}, which were executed with the moving-mesh code AREPO \citep{Springel2010}. With a comprehensive galaxy formation model built into AREPO, IllustrisTNG (hereafter TNG) is capable of realistically tracking the clustering evolution of dark matter and baryons in the universe. The TNG suite includes three simulation volumes: TNG100, TNG300 and TNG50. In this study, we focus on the highest-resolution version of the TNG100 simulation, i.e. TNG100-1, whose box size is  $L_\mathrm{b}=75 \ \mpch$, with a  dark matter mass resolution of $7.5 \times 10^6 M_\odot$ and baryonic mass resolution of $1.4 \times 10^6 M_\odot$. In order to perform the wavelet analysis of density fields at redshift $z=0$, we use cloud-in-cell (CIC) assignment scheme to assign all the mass points to a $1024^3$ uniform mesh, thereby obtaining mass density distribution at Cartesian grids.

%---------------------------------------------------------------------------------
\section{Results}
\label{sec:results}

%---------------------------------------------------------------------------------
\subsection{Wavelet analysis of 1D Zel'dovich density fields}

From Fig.~\ref{fig:zel_dens} we see that characteristics of the 1D density growth in our case is similar to the 3D simulations. With evolution of this simple system from linear ($\delta\ll 1$) to highly non-linear regime ($\delta\gg1$) due to gravitational effects, structures become increasingly significant. For example, there is an obvious underdense region roughly ranging from $\eta\sim 0.3$ to $\eta\sim 0.5$, as well as a large overdense region which is next to it. Since these two distinct local environments in the universe are thought to have different effects on matter clustering \citep{Abbas2005}, we hope that wavelet methods are able to distinguish between them in our 1D toy model. In this section, we present the results for 1D matter clustering based on the CWT analysis.

% ---------------------------------------------------------------------------
\begin{figure}
	\centerline{\includegraphics[width=0.45\textwidth]{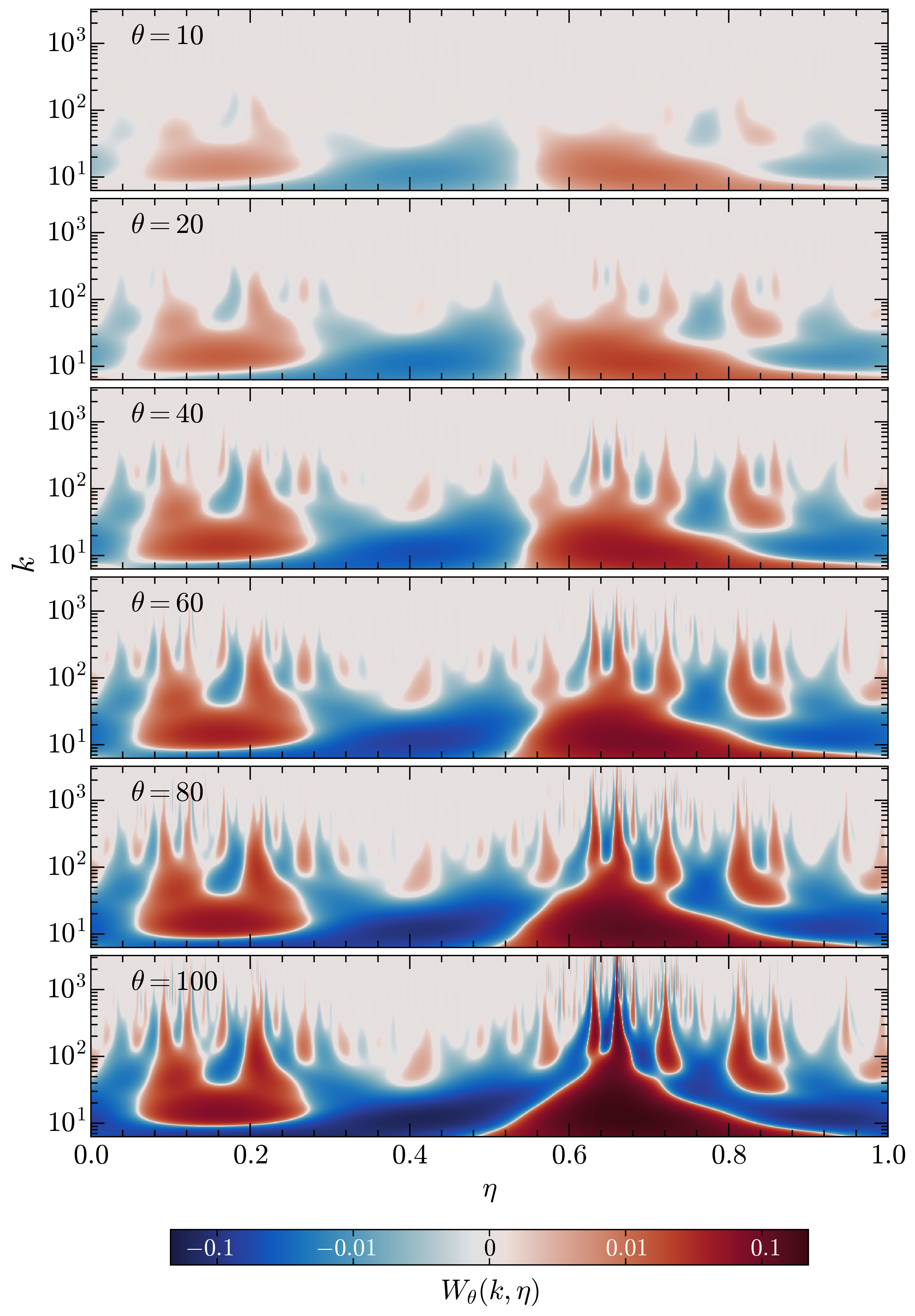}}
	\caption{Wavelet transform plots for Zel'dovich density fields at epochs $\theta=10, 20, 40, 60, 80$ and $100$. The vertical coordinate $k$ represents $\sqrt{5}/2$ times the scale parameter $w$ according to equation~(\ref{eq:scale&wavenum}). These plots show the scale growth and the spatial location where it occurs.}
	\label{fig:cwt_zel}
\end{figure}

%-------------------------------------------------------------------------------
\subsubsection{Space-scale decomposition of the density fields}

As stated by equation~(\ref{eq:CWT}), the CWT is defined for the infinite input signal. Hence the finite signal must be padded with some values before the transform is performed. The usual padding schemes include zero padding, decay padding, periodic padding and symmetric padding \citep{Addison2017}. In the present work, the density contrasts are periodically padded with themselves since they satisfy the periodic boundary condition. Then by employing the CWT, different features are picked out by GDW at each scale while retaining positional information, as illustrated in Fig.~\ref{fig:cwt_zel}. From this figure, we can see that the wavelet coefficients $W_\theta(k, \eta)$ take values from negative (blue) to positive (red), reflecting that the density field is anti-correlated and correlated with GDW, respectively. Positive wavelet coefficients correspond to the location of the density peaks, and those negative coefficients are distributed between the density peaks. This is due to the fact that the shape of the GDW matches well with the density peaks, which is more evident in Fig. \ref{fig:cwt_tng}.

Let's focus on $|W_\theta(k, \eta)|$. The characteristics of matter clustering will be seen qualitatively from the space-scale plane. From $\theta = 10$ to $20$, $|W_\theta(k, \eta)|$ evolves little with time and is dominated by large scale components with a relatively random spatial distribution. This indicates that the density field is almost homogeneous. At $\theta=40$, the small-scale components start to become apparent owing to the gravitational interactions. Since then, $|W_\theta(k, \eta)|$ progressively increases with time. As a consequence, strongly structured patterns are formed in the space-scale plane at $\theta = 100$, suggesting that the density field is highly nonhomogeneous at this time. It is noteworthy that all scale components grow very significantly in the region from $\eta\sim0.5$ to $\eta\sim0.8$. As a contrast, there is almost no small scale components generated between $\eta\sim 0.3$ to $\eta\sim 0.5$, while there is a moderate increase at small scales in other regions. Based on these simple analyses, we find that the CWT of density contrast reproduces its local evolutionary features very well.

% ---------------------------------------------------------------------------
\begin{figure}
	\centerline{\includegraphics[width=0.45\textwidth]{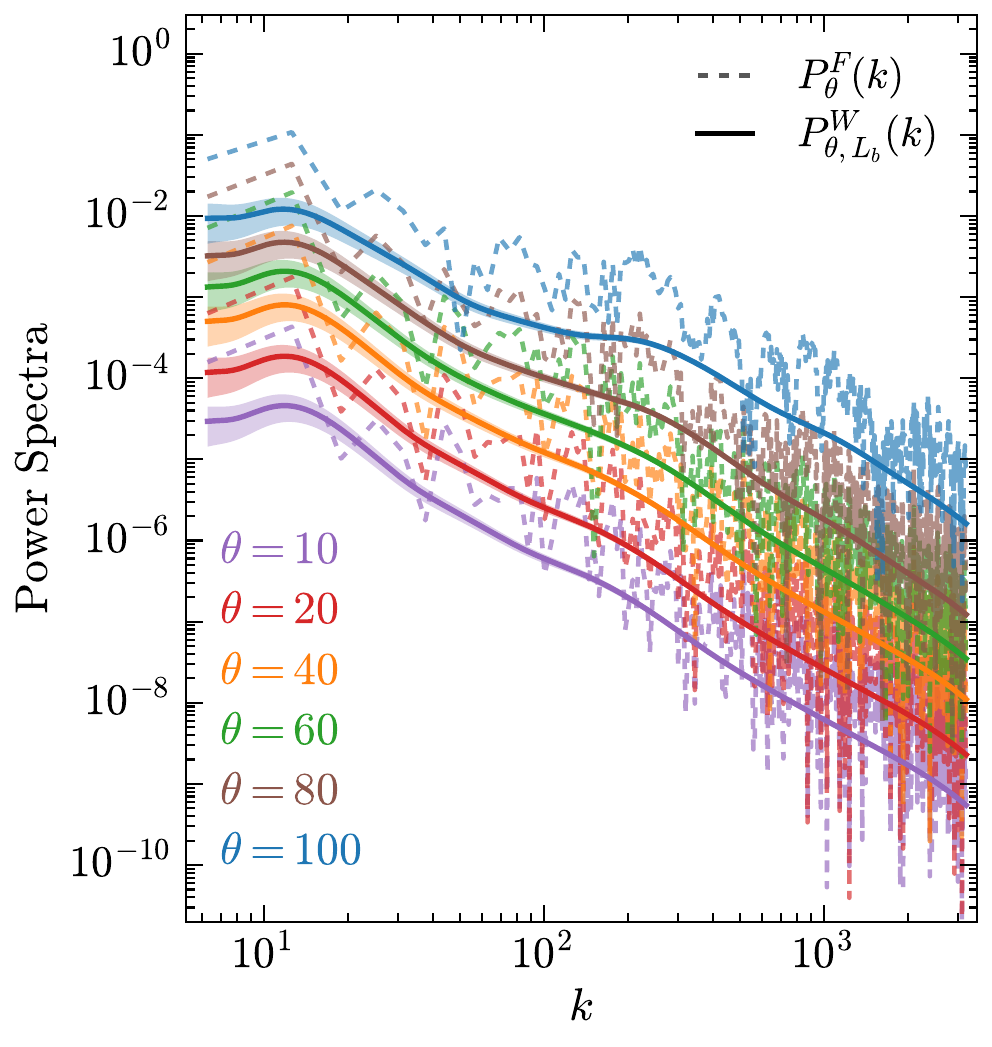}}
	\caption{WPS (solid lines) and Fourier power spectra (dashed lines) of Zel'dovich density fields at epochs $\theta=10, 20, 40, 60, 80$ and $100$, as labeled. The color bands corresponding to each WPS show their statistical errors estimated by equation~(\ref{eq:wp_error}).}
	\label{fig:wps&fps_zel}
\end{figure}

% ---------------------------------------------------------------------------
\begin{figure}
	\centerline{\includegraphics[width=0.48\textwidth]{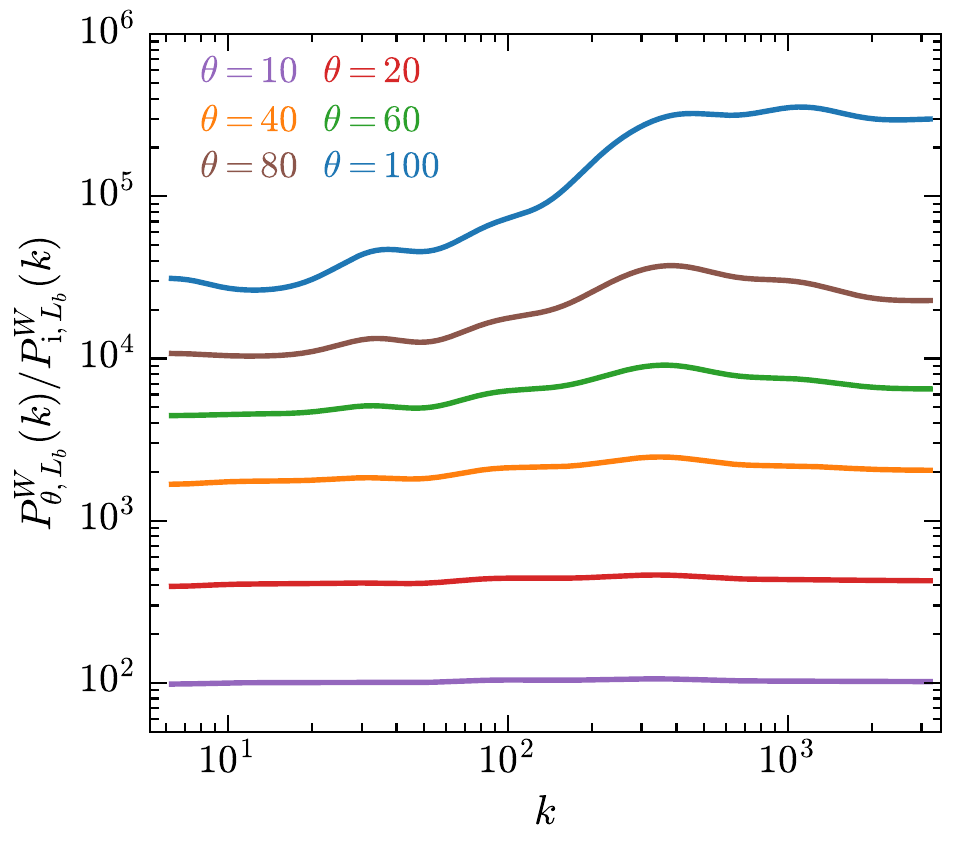}}
	\caption{The ratio of the WPS at each epoch to the initial WPS. We observe that, $P^{W}_{\theta,L_b}(k)/P^{W}_{\mathrm{i},L_b}(k)\approx\theta^2$ at linear stages, while this relation does not hold at late times.}
	\label{fig:rwps_zel}
\end{figure}

%----------------------------------------------------------------------------
\begin{figure}
	\centerline{\includegraphics[width=0.45\textwidth]{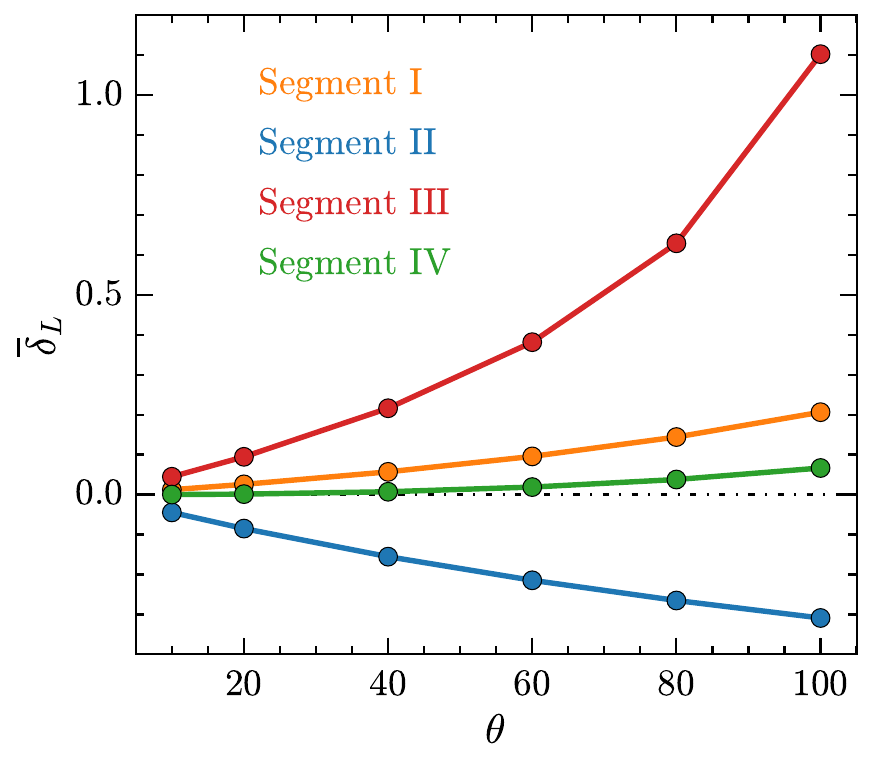}}
	\caption{Time evolution of mean density contrasts within four local regions. Segment III evolves into a highly overdense region, while Segment II evolves into a underdense region. Segements I and IV are slightly overdense regions.}
	\label{fig:mean_local_dens_zel}
\end{figure}

%----------------------------------------------------------------------------
\begin{table*}
	\newcolumntype{M}[1]{>{\centering\arraybackslash}m{#1}}
	\caption{We split the density field at each time into four consecutive segments. The spatial range spanned by each segment and the corresponding local mean density contrast at times of $\theta=10, 20, 40, 60, 80$ and $100$ are listed.}
	\label{tab:segs}
	{\renewcommand{\arraystretch}{1.4}
		\begin{tabular}{M{1cm}  | M{3.5cm}  M{1.2cm}  M{1.2cm}  M{1.2cm}  M{1.2cm}  M{1.5cm}  M{1.5cm}}
			\toprule
			\multirow{2}{*}{Segment} & \multirow{2}{*}{Spatial Range}    & \multicolumn{6}{c}{The mean density contrast $\overline{\delta}_L$} \\
			&    & $\theta$=10 & 20 & 40 & 60 & 80 & 100 \\
			\hline
			I   &$0.00\lesssim\eta\lesssim 0.27$ & 0.012  & 0.025 & 0.056 & 0.095 & 0.144 & 0.206\\ \\[-1.5em]
			II  &$0.27\lesssim\eta\lesssim 0.54$ & -0.045 &-0.085 &-0.155 &-0.214 &-0.265 &-0.309\\ \\[-1.5em]
			III &$0.54\lesssim\eta\lesssim 0.78$ & 0.044  & 0.094 & 0.216 & 0.381 & 0.629 & 1.102\\ \\[-1.5em]
			IV  &$0.78\lesssim\eta\lesssim 1.00$ & 0.000  & 0.001 & 0.007 & 0.018 & 0.037 & 0.066\\
			\hline
	\end{tabular}}
\end{table*}

%----------------------------------------------------------------------------
\begin{figure*}
	\centerline{\includegraphics[width=0.9\textwidth]{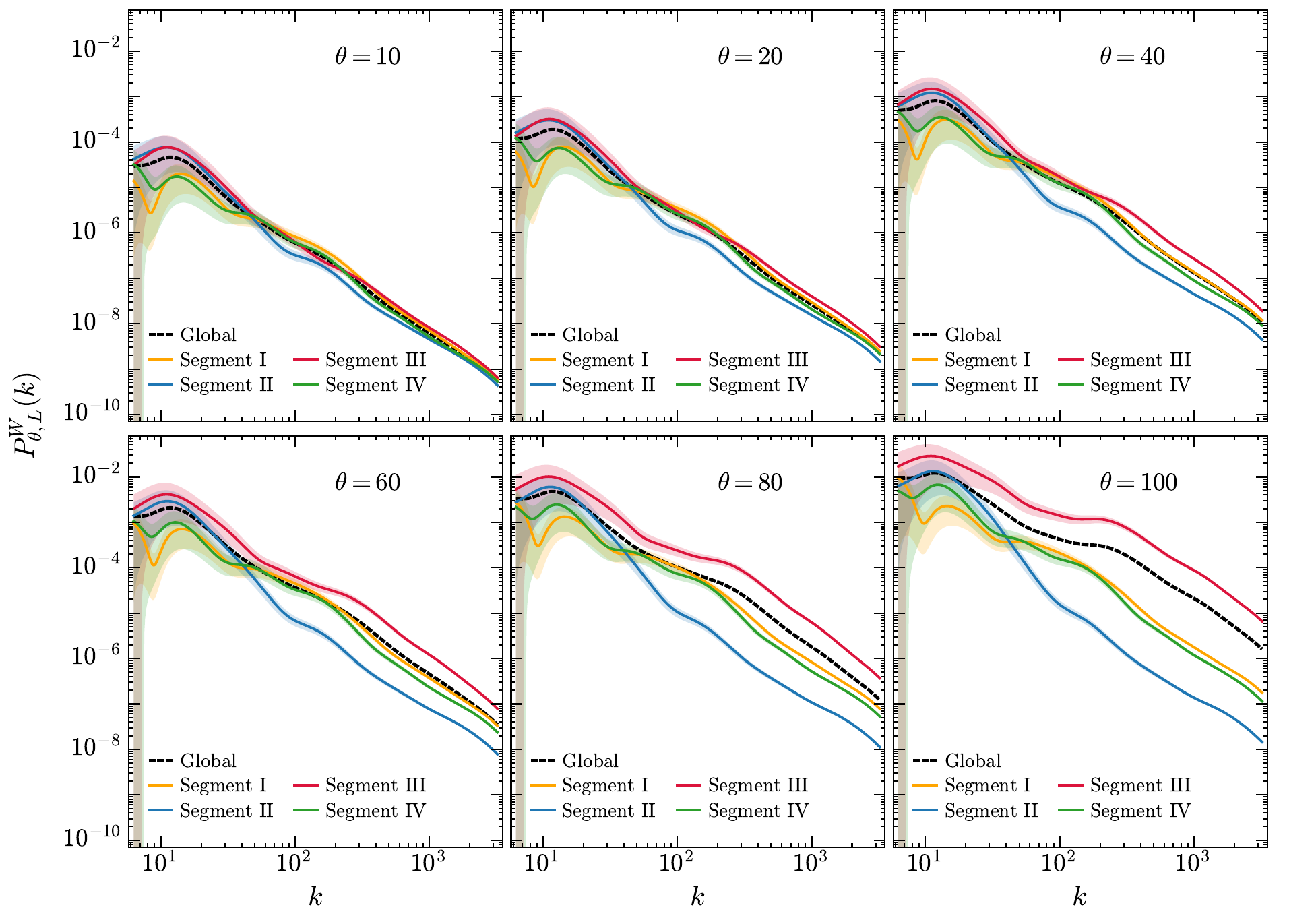}}
	\caption{Local WPS for different density environments at times $\theta=10, 20, 40, 60, 80$ and $100$. The shaded area for each curve represents the statistical error, from which we can see that those local WPS are less significant statistically on scales of $k\lesssim 40$. }
	\label{fig:local_wps_zel}
\end{figure*}

%------------------------------------------------------------------------------------------------
\begin{figure*}
	\centerline{\includegraphics[width=0.9\textwidth]{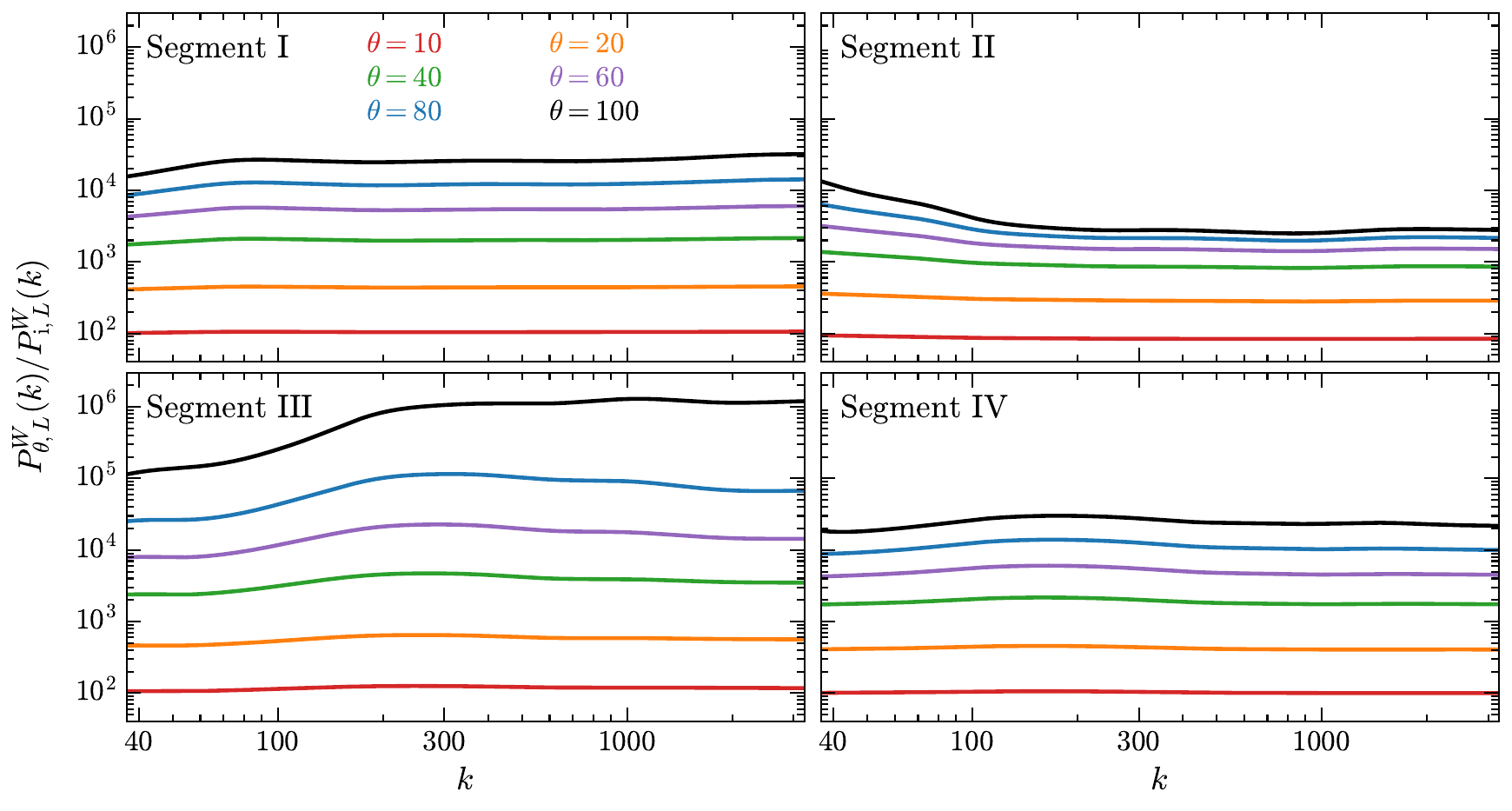}}
	\caption{Local WPS for different density environments at times $\theta=10, 20, 40, 60, 80$ and $100$ relative to the initial WPS.}
	\label{fig:local_rwps_zel}
\end{figure*}
%-------------------------------------------------------------------------------------------------
\begin{figure}[t]
	\centerline{\includegraphics[width=0.5\textwidth]{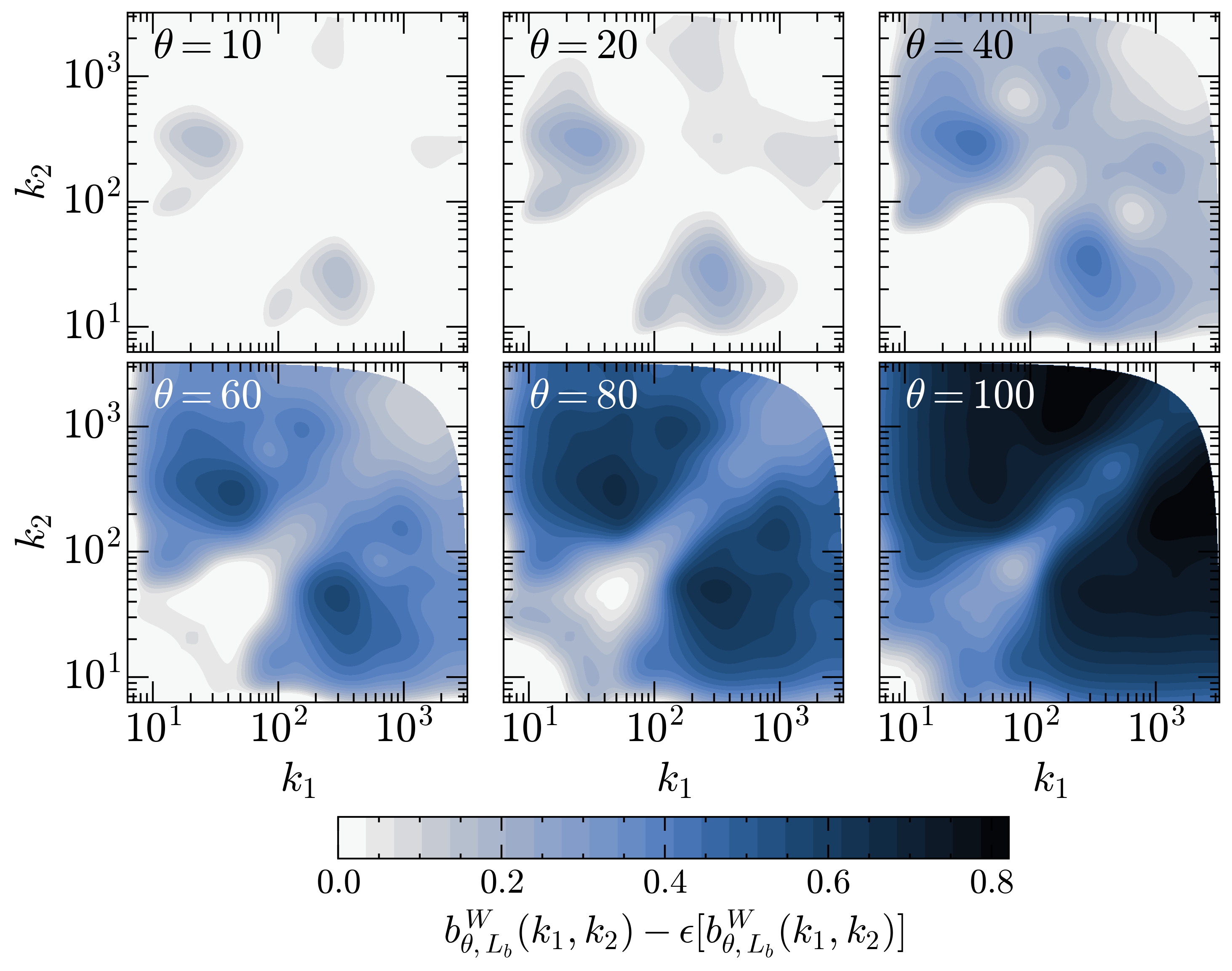}}
	\caption{Contour plots of the WBCs with statistical noise level subtracted for density fields at times $\theta=10, 20, 40, 60, 80$ and $100$. In these plots, values of the WBCs less than the statistical noise level are set to be zero.}
	\label{fig:global_wb_zel}
\end{figure}
%-------------------------------------------------------------------------------------------------
\begin{figure}
	\centerline{\includegraphics[width=0.48\textwidth]{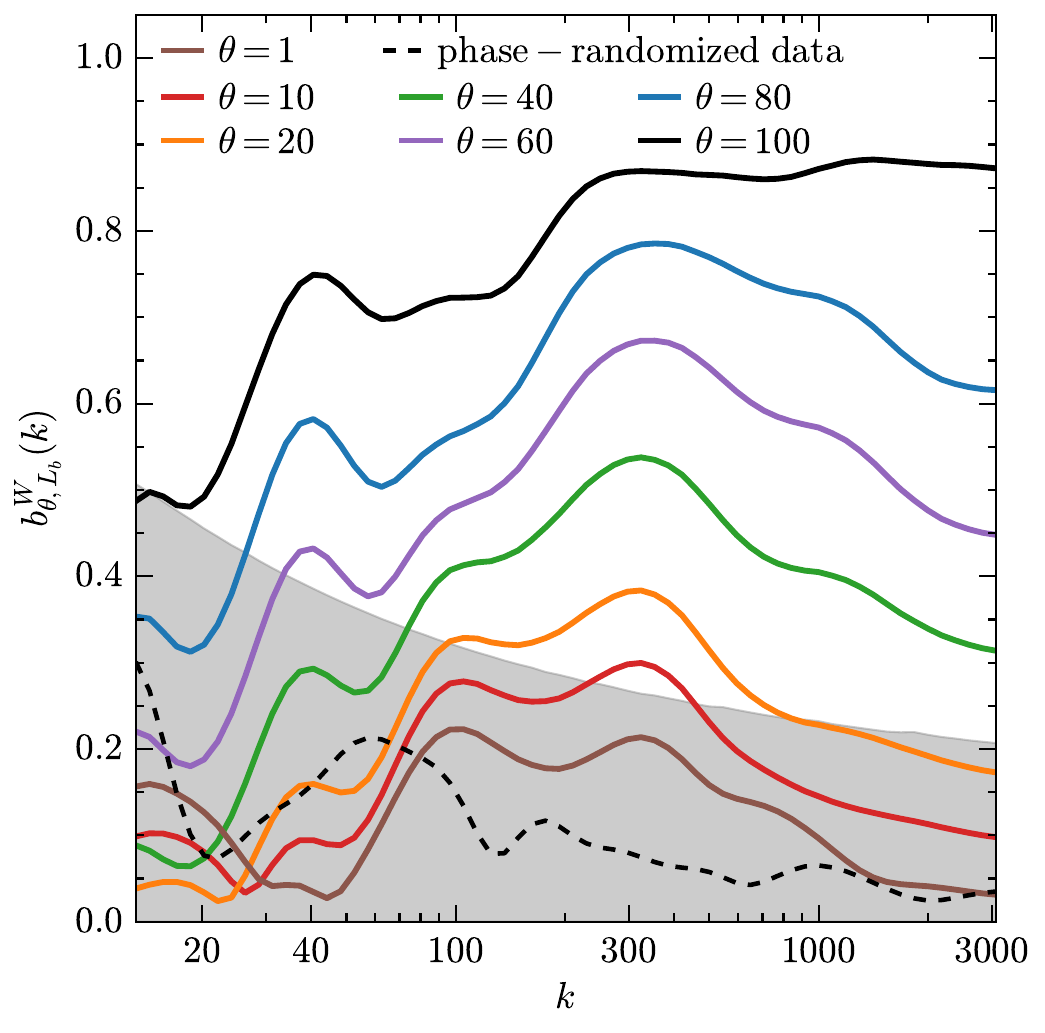}}
	\caption{Summed WBC for density fields at times $\theta=1, 10, 20, 40, 60, 80$ and $100$ (solid lines). The summed WBC for the phase-randomized data obtained from the density field at $\theta=100$ is indicated by the black dashed line and that of initial density field ($\theta=1$) is indicated by the brown solid line as a comparison. The gray area represents the region where the summed WBC is less than the statistical noise level.}
	\label{fig:summed_wb_zel}
\end{figure}

%----------------------------------------------------------------------------
\begin{figure}
	\centerline{\includegraphics[width=0.48\textwidth]{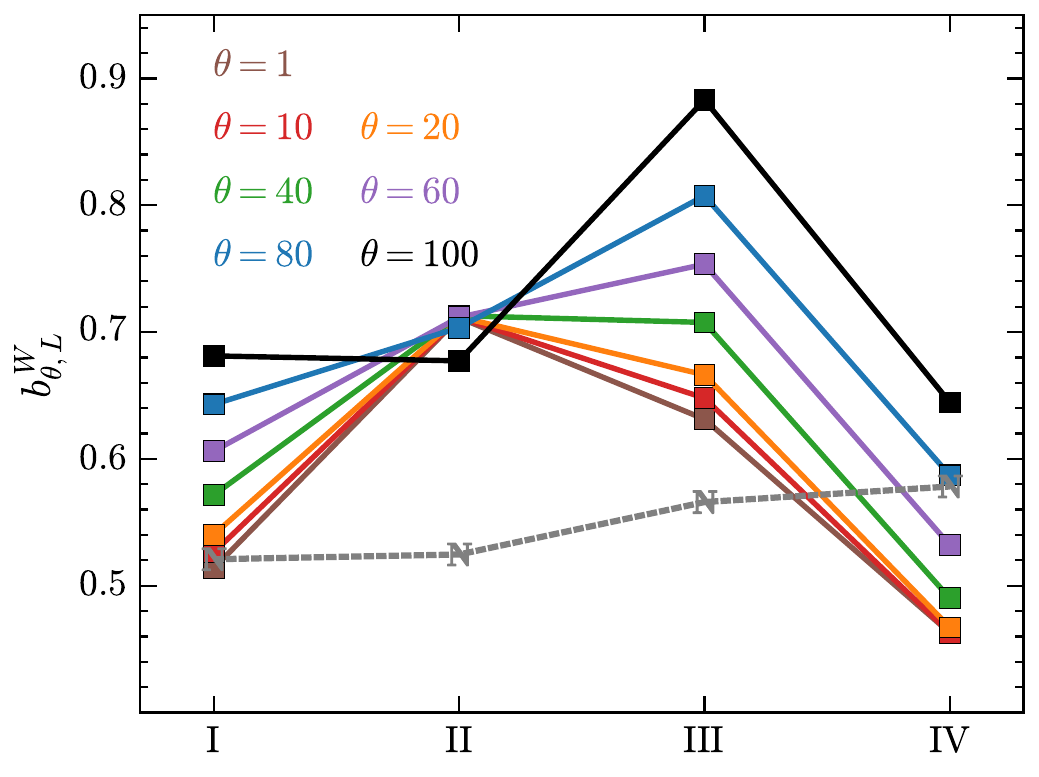}}
	\caption{Time evolution of the total WBC from the initial time $\theta=1$ to $\theta=100$ in Segments I, II, III and IV. The marker `N' indicates the noise level for each segment.}
	\label{fig:local_total_wbc_zel}
\end{figure}
%----------------------------------------------------------------------------
\begin{figure*}
	\centerline{\includegraphics[width=0.85\textwidth]{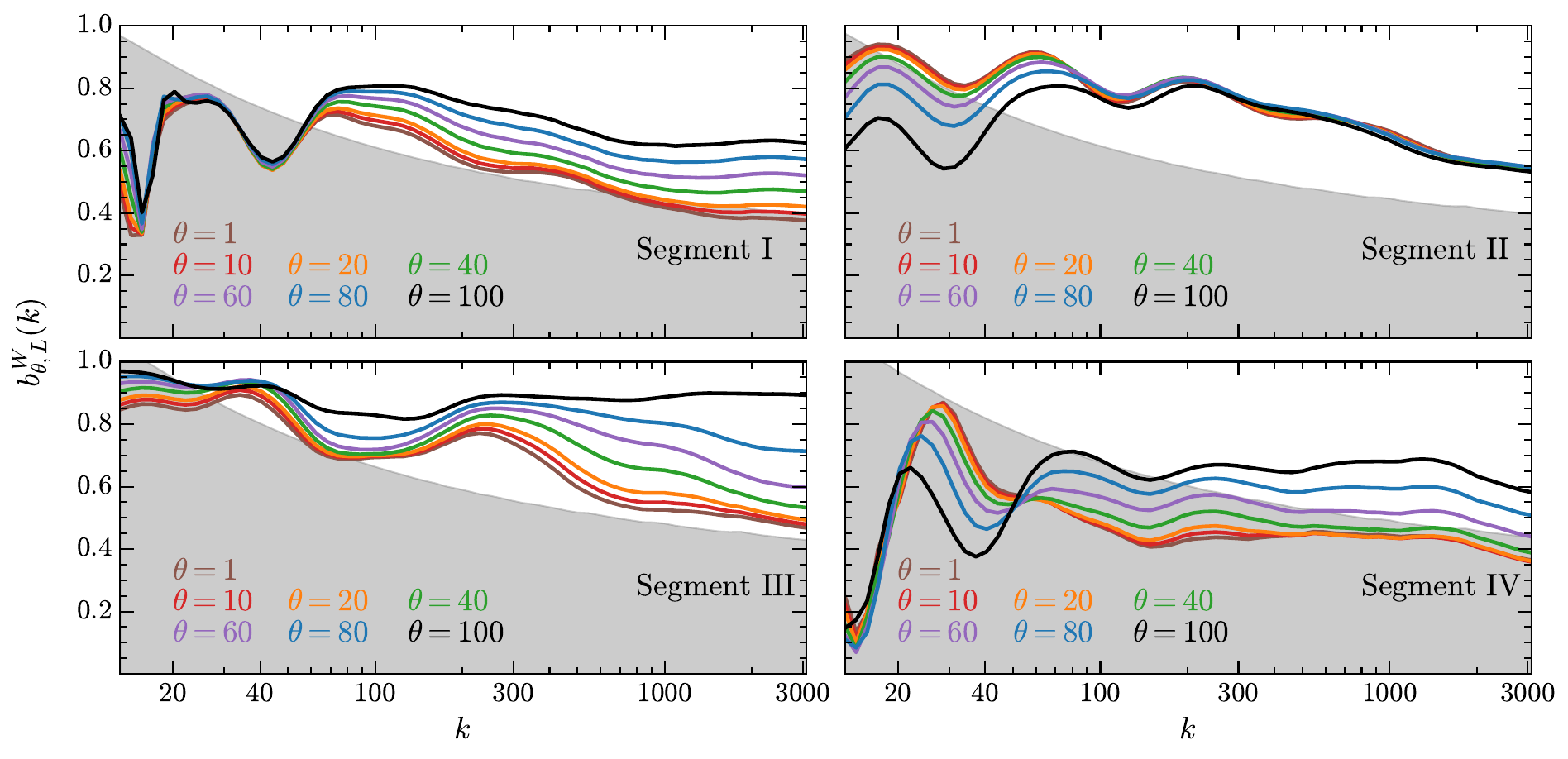}}
	\caption{The summed WBC of four consecutive segments marked I, II, III and IV at times $\theta=1, 10, 20, 40, 60, 80$ and $100$. The grey region of each subplot indicates where the summed WBC are below the statistical noise level. }
	\label{fig:local_swbc_zel}
\end{figure*}

%----------------------------------------------------------------------------
\begin{figure*}
	\centerline{\includegraphics[width=0.85\textwidth]{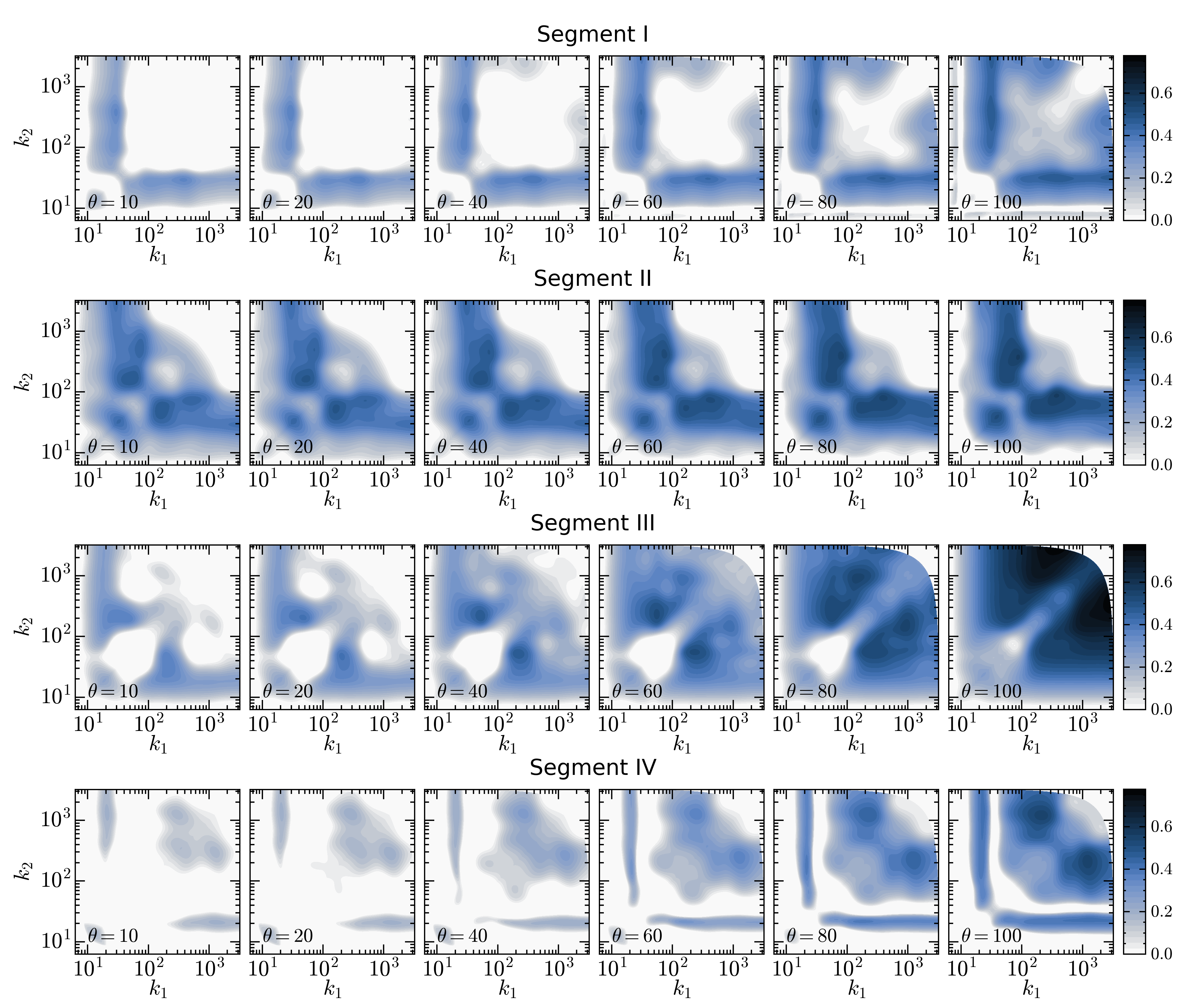}}
	\caption{WBC with statistical noise subtracted, i.e. $b^{W}_{
			\theta,L}(k_1,k_2) - \epsilon[b^{W}_{\theta,L}(k_1,k_2)]$ for local segments at times $\theta = 10, 20, 40, 60, 80$ and $100$. For easier visual description, values of the WBC less than the statistical noise level are set to be zero in each subplot.}
	\label{fig:local_wbc_zel}
\end{figure*}

%------------------------------------------------------------------------------
\subsubsection{Wavelet power spectrum measurements}

By averaging squared wavelet coefficients over a local spatial region, we thus obtain the local WPS. For a better understanding of the WPS, we first compare the global wavelet and Fourier power spectrum of 1D Zel'dovich fields. A main visible feature in Fig.~\ref{fig:wps&fps_zel} is that the WPS's are smoother than their Fourier counterparts, as expected by equation~(\ref{eq:wps&fps}). These two types of power spectra have approximately same trends with comparable amplitudes. In fact, the global WPS is able to reproduce the correct exponent of the power-law Fourier power spectrum \citep{Wang2021}, although it is not obvious for our 1D noisy data. From the WPS in Fig.~\ref{fig:wps&fps_zel}, we see that the strength of clustering on small scales increases with time. Furthermore, by measuring the change of the WPS at each time relative to the initial WPS, it should help better demonstrate the nonlinear evolution of clustering. Since $\delta(\eta,t)$ approximately equals to $\theta(t)F(\eta)$ in the linear regime, there is a relation between the WPS of $\delta(\eta,t)$ and the initial one given by
%------------------------------------------------------------------------
\begin{flalign}
\label{eq:wps&initial_wps}
P^{W}_{\theta,L_b}(k)\approx \theta^2 P^{W}_{\mathrm{i},L_b}(k).&&
\end{flalign}
Hence if the ratio of these two WPS, $P^{W}_{\theta,L_b}(k)/P^{W}_{\mathrm{i},L_b}(k)$, is not scale dependent, then the growth of the density fluctuations is linear. We show this ratio for times $\theta=10, 20, 40, 60, 80$ and $100$ in Fig.~\ref{fig:rwps_zel}. Obviously, $P^{W}_{\theta,L_b}(k)/P^{W}_{\mathrm{i},L_b}(k)$ is almost constant over the entire scale range at $\theta=10$ and $20$, suggesting that the density contrast is linear. However, this ratio becomes scale dependent at the time $\theta=40$ and there is a mild bump around $k\approx 300$, which means that there are a few newly generated scale components and the density contrast is quasi-nonlinear. Since then, this bump has been enhanced by nonlinear effects with time to $\theta=80$, while the ratio remains approximately flat on large scales of $k\lesssim 20$. At the time $\theta=100$, the ratio increases more significantly on small scales thus leading to a plateau at scales of $k\gtrsim300$, implying that the density field has developed into the highly non-linear regime.

Next, we turn our attention to the environmental dependence of matter clustering. To do this, we split the density field into four consecutive segments labeled I, II, III and IV. The spatial extent spanned by each segment and the mean density within that local space at each time are listed in Table~\ref{tab:segs}. The evolutionary trends of these four local mean densities are more clearly depicted in Fig.~\ref{fig:mean_local_dens_zel}. We see that Segment III represents the highly overdense environment, and both Segments I and IV represent the slightly overdense environment, while Segment II represents the underdense environment. These different density environments are expected to exhibit distinct clustering characteristics from each other. Therefore, it is instructive to examine the local WPS for each environment, as shown in Fig.~\ref{fig:local_wps_zel}. Each subplot shows the WPS for different regions at the same time. At linear stages ($\theta=10$ and $20$), all of these local power spectra roughly converge to the global WPS. Starting from $\theta = 40$, however, the divergence between these local power spectra becomes more and more significant with time. For the highly overdense environment, i.e. Segment III, its WPS is more enhanced than all other regions over the entire scale range. In contrast, the WPS for the underdense environment (Segment II) is more suppressed than all other regions on scales of $k\gtrsim 40$. For Segments I and IV, as slightly overdense regions, the amplitudes of their WPS are very close to each other and fall between II and III on the scales of $k\gtrsim 40$. Notice that as the scale becomes larger, the statistical error in the power spectrum becomes increasingly large to the extent that it does not give a significant description of matter clustering in the range $k \lesssim 40$, due to the wavelet coefficients being non-orthogonal and the length of local region being too short. Even so, the WPS is still able to give a meaningful interpretation over a relatively large scale range.

In Fig.~\ref{fig:local_rwps_zel}, we also measure the local WPS at each time relative to its corresponding initial WPS within the scale range $k\gtrsim 40$ where statistical errors are smaller. For all segments, we see that $P^{W}_{\theta,L}(k)/P^{W}_{\mathrm{i},L}(k)$ approximates to $\theta^2$ at stages of $\theta=10$ and $20$. However, at late stages, the evolutionary trends of Segments II and III  are completely opposite. For Segment II, the growth of small-scale components is getting slower, but for Segment III, the small-scale components are growing faster and faster. Although the WPS's of Segments I and IV are very similar, there are visible differences between them. Specifically, the former has a slightly upward tilt on small scales, while the latter remains roughly horizontal.

Based on these facts, we find that WPS is fully capable of detecting the effects of density environments on matter clustering. However, as pointed out above, the density field evolves into the non-linear regime at late times, indicating that there are couplings between different scale components. The WPS, as two-point statistics, cannot determine such scale-coupling since it does not contain phase information. To detect scale-coupling, we measure the WBC, which is the lowest order statistics sensitive to nonlinear couplings between different scales of matter density field.

%---------------------------------------------------------------------------------
\subsubsection{Wavelet bicoherence measurements}

Fig.~\ref{fig:global_wb_zel} shows the time evolution of matter WBC with statistical noise subtracted, i.e. $b^{W}_{\theta,L_b}(k_1,k_2)-\epsilon[b^{W}_{\theta,L_b}(k_1,k_2)]$. Note that $b^{W}_{\theta,L_b}(k_1,k_2)$ is symmetric about the diagonal where $k_1=k_2$, as indicated by equations~(\ref{eq:wbs}) and (\ref{eq:wbc}). So we will only concern the part below (or above) this diagonal. For visual convenience, values of bicoherence less than the noise level are not considered, because they are less significant physically. To better understand the statistical noise level, as suggested by \citet{Milligen1995a, Milligen1995b}, we perform the FFT on the highly non-linear density field at time $\theta=100$, and give each Fourier component a random phase while maintaining its amplitude, then we perform the inverse FFT to get a new set of data. Such new phase-randomized density field is expected to have no structures induced by scale-coupling, while its power spectrum is identical to that of the raw density field, in which structures are well formed, as is illustrated in Fig.~\ref{fig:summed_wb_zel}. It can be seen that the summed WBC of the density field at $\theta=100$ is much higher than that of the phase-randomized version, which falls below the noise level along with the initial density field. Thus the statistical noise level provides us a criterion to discriminate matter distributions with structures formed from those without. Accordingly, looking at Fig.~\ref{fig:global_wb_zel} and \ref{fig:summed_wb_zel} together, there is a very weak coupling between large scales around $k\sim 10$ and intermediate scales around $k \sim 300$ at times $\theta=10$ and $20$. Gradually, the scale-coupling becomes more and more significant with time and eventually spreads to the whole scale range, which implies that the matter distribution becomes more and more structured.

In Fig.~\ref{fig:local_total_wbc_zel}, we consider the evolution of the total WBC with time in the consecutive segments I, II, III and IV. An interesting phenomenon is that the degree of non-linearity in Segment II, the underdense region, is much higher than noise level at the initial time, and this non-linearity is less evolved. On the other hand, the non-linearity is more evolved in overdense regions, although it is very weak at the initial time. In particular, Segment III, as a highly overdense region, has the strongest non-linearity at late times. This implies that structure formation occurs mainly in Segment III. Furthermore, we examine the scale-coupling for these segments by measuring their WBC and summed WBC at different times and corresponding results are shown in Figs.~\ref{fig:local_swbc_zel} and \ref{fig:local_wbc_zel}. Consistent with the results in Fig.~\ref{fig:local_total_wbc_zel}, the estimated WBC and summed WBC in Segment II exhibit significant values between widely separated scales at the initial time, because the matter distribution within such a underdense region is left-skewed. Both WBC and summed WBC in Segment II barely evolve with time, indicating that there is almost no structure formation in this region. Also, Segment III shows a notable coupling between scale bands $20\lesssim k\lesssim 100$ and $100\lesssim k\lesssim 500$ at initial stages, due to the matter distribution in such a highly overdense region being right-skewed. Unlike Segment II, the WBC and summed WBC of Segment III evolves the most dramatically, which indicates that the matter becomes increasingly clustered. For Segments I and IV, the bicoherences are below or close to the statistical noise level at initial stages and evolve mildly with time, which means that the matter experiences moderate structure formation in these two slightly overdense regions.

%----------------------------------------------------------------------------
\begin{figure*}[h]
	\centerline{\includegraphics[width=0.98\textwidth]{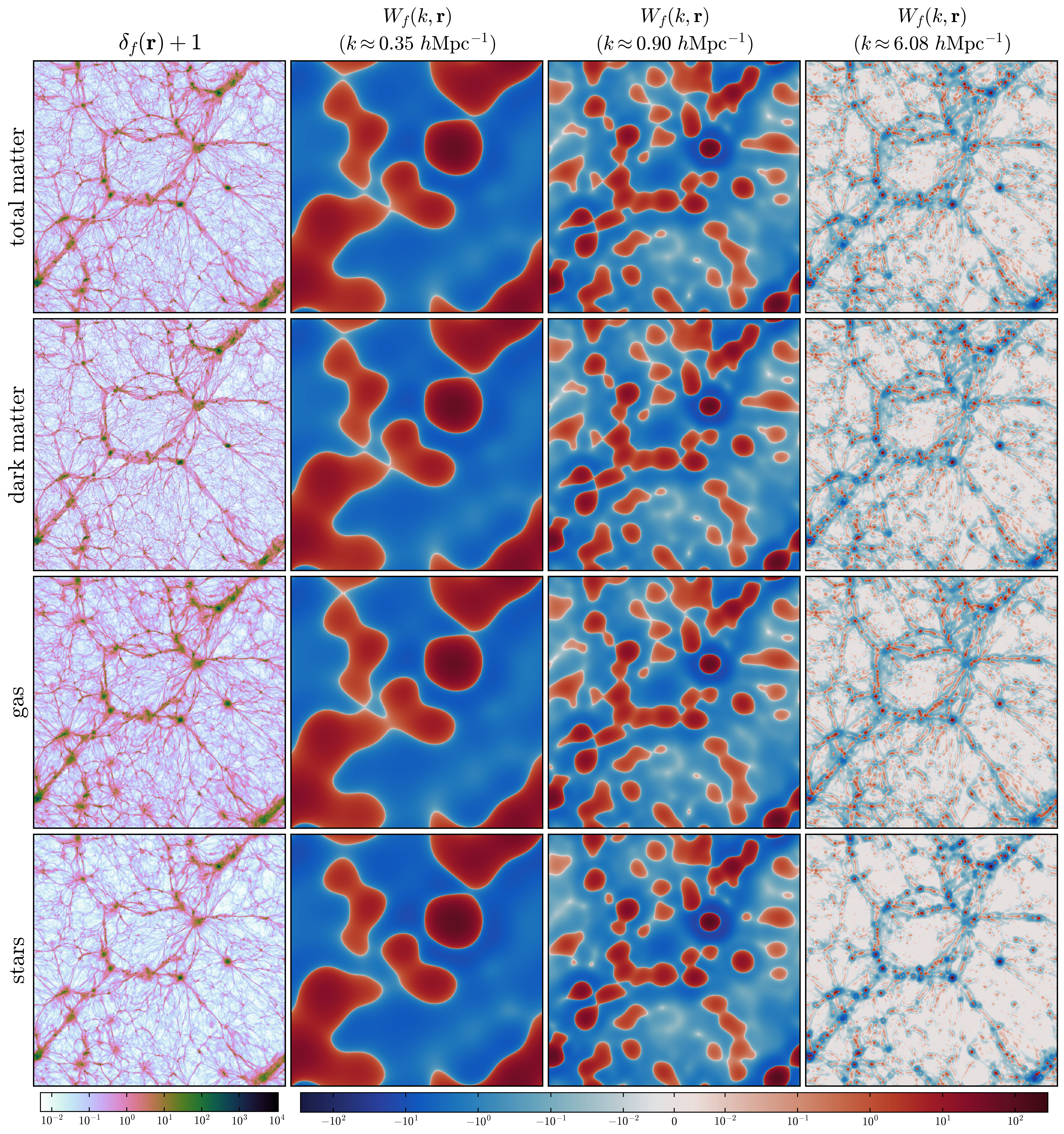}}
	\caption{2D slices of density fields $\delta_f(\mathbf{r})+1$ and their wavelet transform $W_f(k,\mathbf{r})$, where $f$ stands for the total matter, dark matter, gas and stars from top to down respectively. These slices are $75\times75\ h^{-2}\mathrm{Mpc}^2$, with a thickness $73.242 \ h^{-1}\mathrm{kpc}$ (one cell scale). Left panels show the density fields at redshift $z=0$, and next panels are their wavelet coefficients at scales $k\approx0.35$, $0.90$ and $6.08 \ h\mathrm{Mpc}^{-1}$, respectively. Notice that the wavelet scale $w$ is replaced by its equivalent Fourier wavenumber $k$ according to the relation $w=2k/\sqrt{7}$, which is obtained following the method of Section \ref{sec:scale&wavenum}. }
	\label{fig:cwt_tng}
\end{figure*}

%----------------------------------------------------------------------------
\begin{figure}[t]
	\centerline{\includegraphics[width=0.48\textwidth]{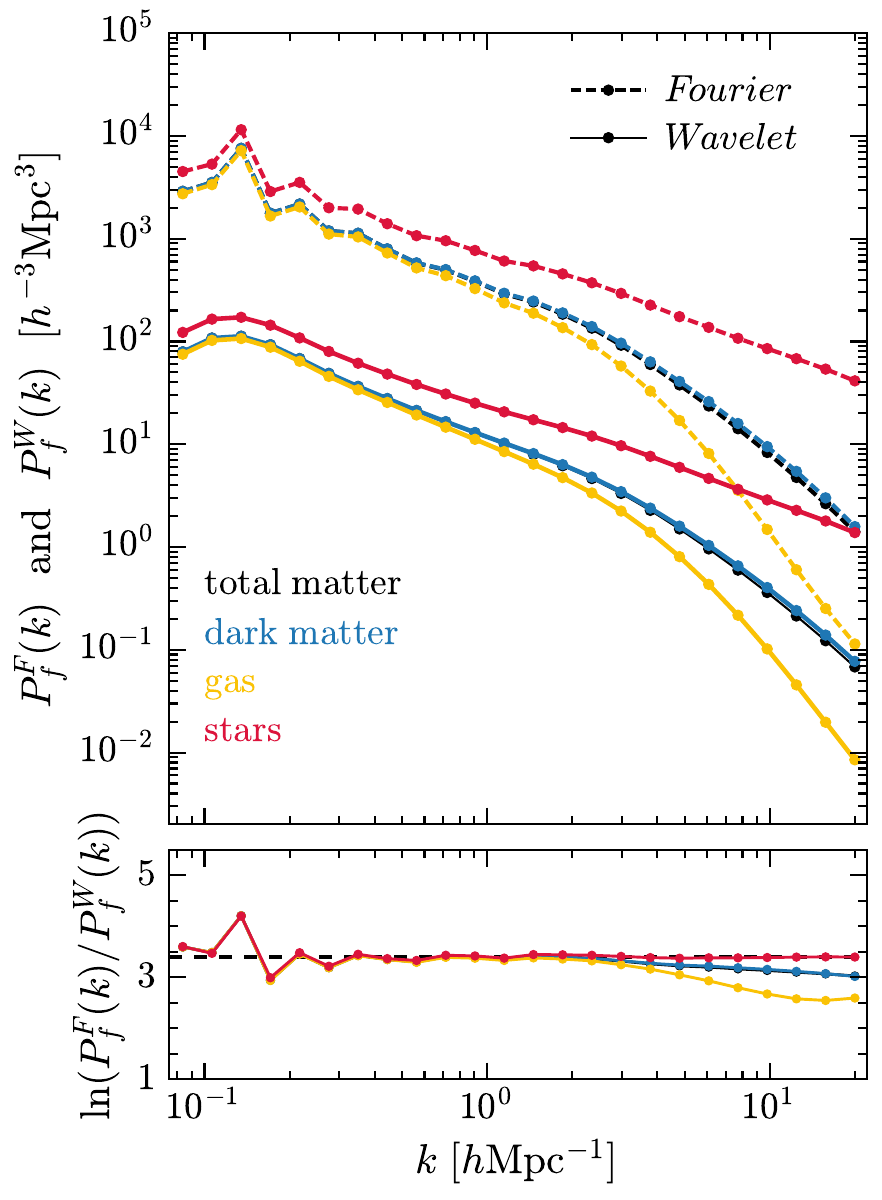}}
	\caption{Top panel: the global wavelet and Fourier power spectra for different matter components at redshift $z = 0$. We show results for the total matter, dark matter, gas and stars, as labelled. Bottom panel: the ratio between the global WPS and Fourier power spectrum for each matter component. The horizontal dashed line denotes the constant $\ln(8\pi^2/\sqrt{7})\approx 3.396$ given by equation \eqref{eq:relation_WPS_FPS}.}
	\label{fig:globalWPS_tng}
\end{figure}

%----------------------------------------------------------------------------
\begin{figure}[th]
	\centerline{\includegraphics[width=0.5\textwidth]{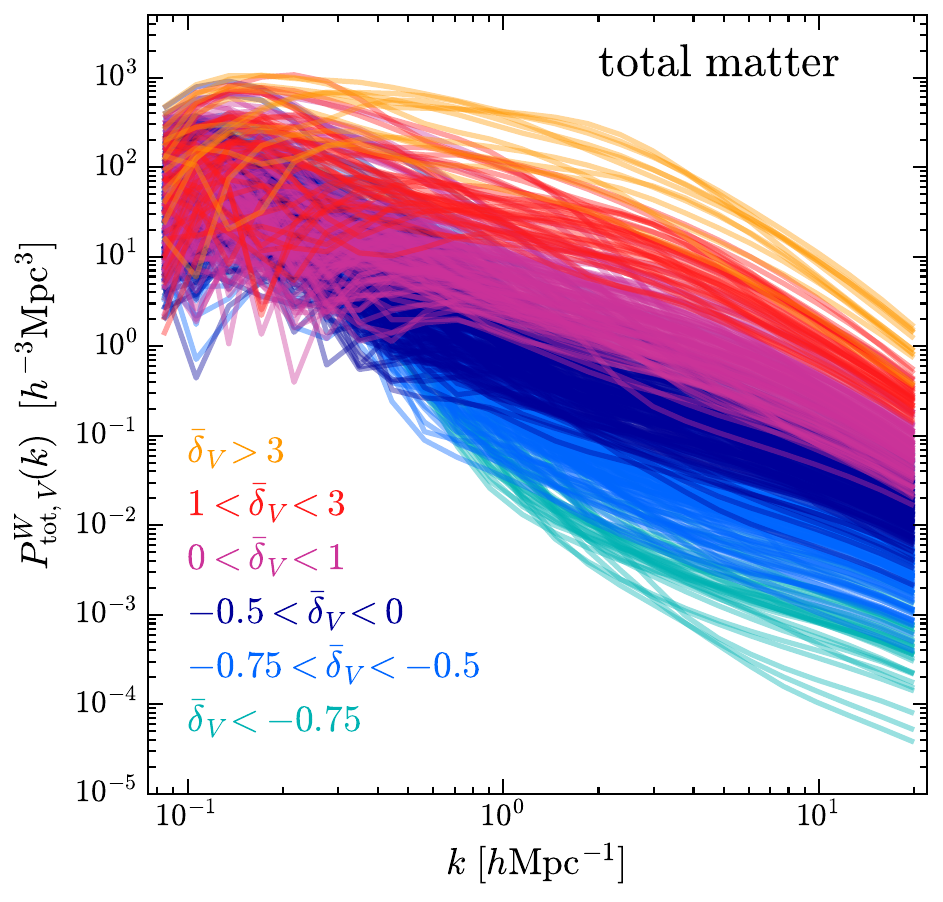}}
	\caption{Local WPSs of the total matter $P^W_{\mathrm{tot},V}(k)$ measured from $N_{sub}^3 = 512$ sub-volumes of $V=9.375^3 \ h^{-3}\mathrm{Mpc}^3$. The color represents the local mean density $\bar\delta_V$ of each sub-volume.}
	\label{fig:localWPS_tot_tng}
\end{figure}

%----------------------------------------------------------------------------
\begin{figure}[th]
	\centerline{\includegraphics[width=0.48\textwidth]{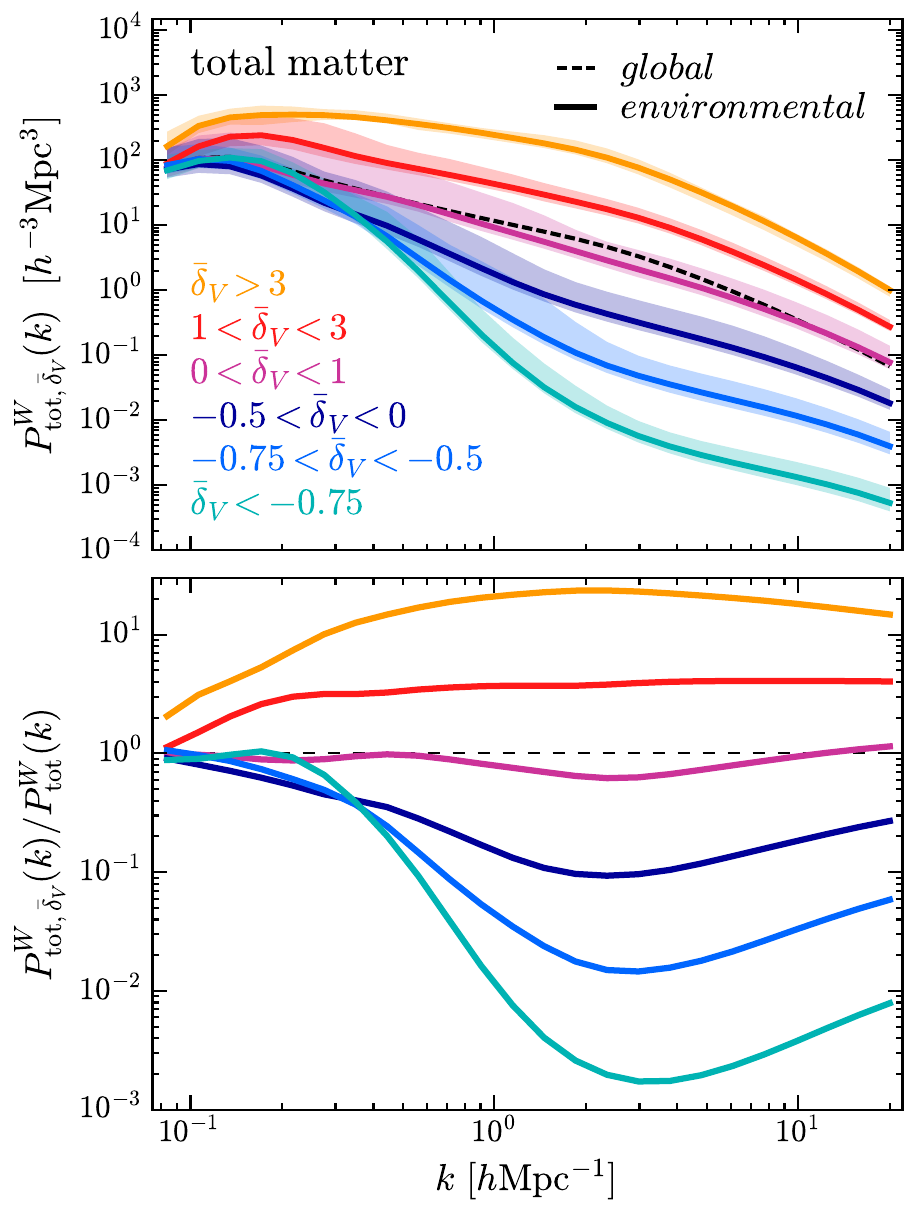}}
	\caption{Top panel: the mean of local WPSs in each density interval for the total matter distribution. The shaded regions represent the standard deviations for the values above and below the mean calculated separately. As a comparison, the global WPS of the total matter distribution is indicated by the dashed line. Bottom panel: the ratio of the mean local WPS in each density interval relative to the global WPS. }
	\label{fig:env_WPS_tng}
\end{figure}
%----------------------------------------------------------------------------
\begin{figure*}[th]
	\centerline{\includegraphics[width=0.99\textwidth]{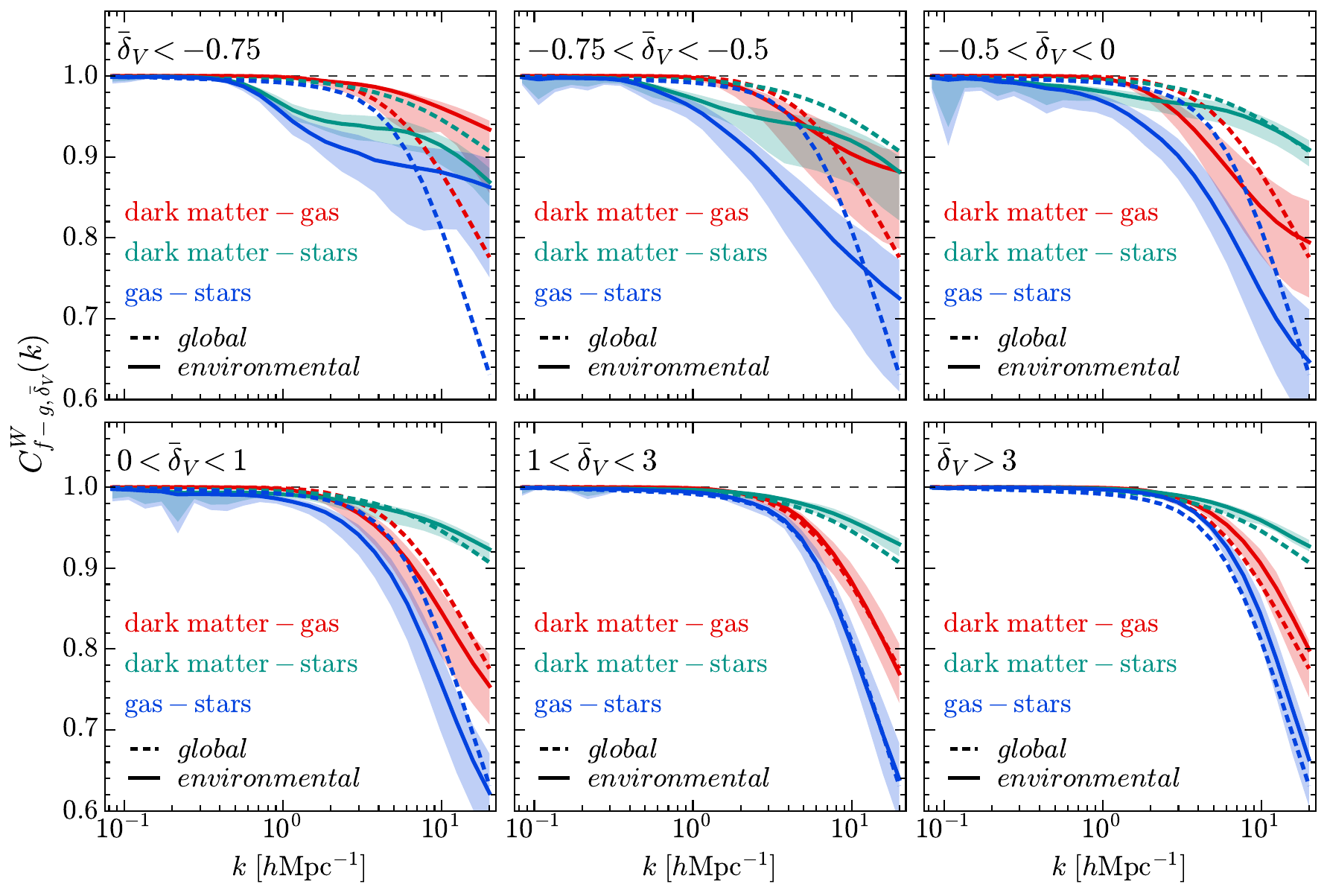}}
	\caption{The mean of local WCCs in different density intervals $C^W_{f-g,\bar\delta_V}(k)$, where $f$ and $g$ stand for dark matter, gas or stars, respectively. The color areas represent the standard deviations for the values above and below the mean calculated separately. The global WCCs are indicated by the dashed lines for reference.}
	\label{fig:env_WCC_tng}
\end{figure*}
%----------------------------------------------------------------------------
\begin{figure*}[th]
	\centerline{\includegraphics[width=0.99\textwidth]{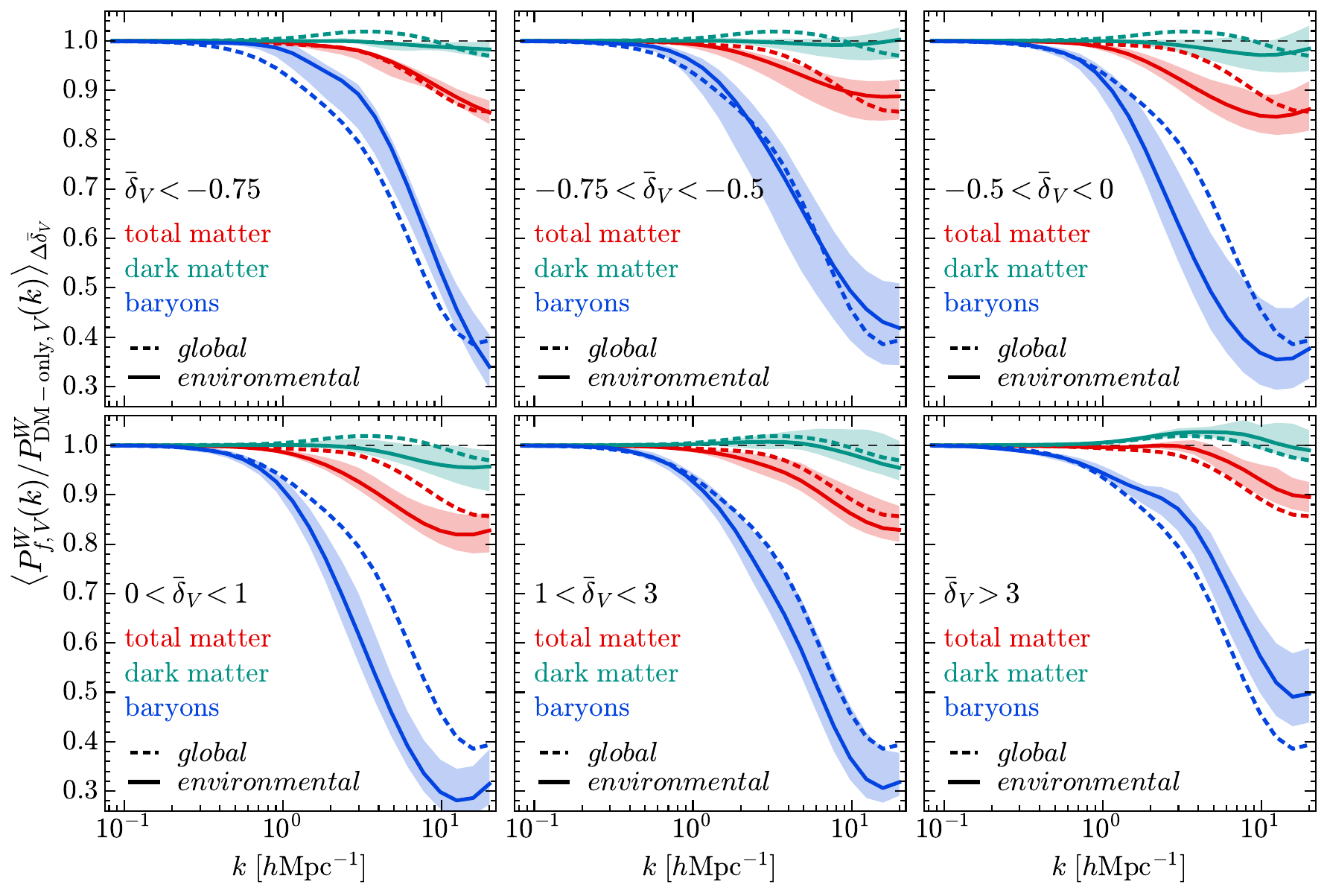}}
	\caption{The mean of the ratio between the local WPSs of the total matter, dark matter, baryons (gas+stars) from the TNG100-1 and the local WPS from the corresponding dark matter-only simulation TNG100-1-Dark in different density intervals. The color areas represent the standard deviations for the values above and below the mean calculated separately. The ratio between the global WPSs are indicated by the dashed lines for reference.}
	\label{fig:env_rWPS_tng}
\end{figure*}
%-------------------------------------------------------------------------------

\subsection{Wavelet analysis of density fields of the TNG100 simulation}

In previous sections, we focus on the wavelet analysis of the non-linear evolution of the 1D Zel'dovich density field, revealing considerable potential of the CWT for characterizing the matter clustering. Therefore, we further apply the 3D continuous wavelet method to the analysis of the TNG100 simulation. For simplicity, we consider the 3D isotropic CWT methods based on the GDW described in Appendix \ref{sec:iso_CWT}, which is the simplest extension of the 1D case. We first perform the isotropic CWT of the 3D density fields of the total matter, dark matter, gas and stars\footnote{Similar to what \citet{Springel2018} did in measuring the cross-correlation functions between different matter, we here put the black holes and stars together, and then labeled as stars.} at redshift $z=0$. In order to avoid the aliasing, we take the maximum scale as $k_\mathrm{max} = k_\mathrm{Nyq}/2$, where $k_\mathrm{Nyq} = 1024 \pi/L_\mathrm{box}$ is the Nyquist frequency of the density grid. For the purpose of presentation, wavelet transforms of density slices at three scales are shown in Fig.~\ref{fig:cwt_tng}. By visual inspection, We see that for all matter components, only large voids and filaments are captured at large scale of $k\approx 0.35 \ h\mathrm{Mpc}^{-1}$, whereas as the scale becomes smaller, the wavelet transform is increasingly accurate in tracking the finer web-like structure of the universe. More quantitative statements can be made by using statistics formulated on the wavelet transform.

In Fig.~\ref{fig:globalWPS_tng}, we compare the global wavelet and Fourier power spectra for different matter distributions. It can be seen that these two types of spectra for each matter have a similar shape. The global WPS of the total matter falls slightly below that of the dark matter at small scales. The gas becomes significantly less clustered than dark matter at intermediate and small scales, while the stellar mass shows a very strong clustering. However, the magnitude of the wavelet power spectrum is different from the Fourier power spectrum. More precisely, the deviations between these two spectra are measured by the ratio between them, which appears to be roughly a constant determined by equation \eqref{eq:relation_WPS_FPS}.

To study the impact of large-scale environment on the matter clustering, we divide the whole simulation volume $V_b=L_b^3$ into $N_{sub}^3$ sub-volumes of $V=(L_b/N_{sub})^3$, where $N_{sub}$ is set to be $8$ and hence the side length $L_b/N_{sub}$ is $9.375 \ h^{-1}\mathrm{Mpc}$. In Fig. \ref{fig:localWPS_tot_tng}, we show the local WPSs of the total matter spatial distribution measured in the sub-volumes. These local WPSs are separated into six categories, according to the mean densities of sub-volumes as follows, 
\begin{enumerate}[label=(\roman*),align=left,leftmargin=*]
\item $\bar\delta_V>3$,
\item $1<\bar\delta_V<3$,
\item $0<\bar\delta_V<1$,
\item $-0.5<\bar\delta_V<0$,
\item $-0.75<\bar\delta_V<-0.5$,
\item $-1<\bar\delta_V<-0.75$,
\end{enumerate}
where $\bar\delta_V=\langle \delta_\mathrm{tot}(\mathbf{r})\rangle_V$ denotes the local mean density of the total matter within a sub-volume. We see that the local WPSs are greater for sub-volumes with higher local mean density for scales up to a few times $0.1 \ h\mathrm{Mpc}^{-1}$. To make it more explicit, we consider the mean of the local WPSs over all sub-volumes within each density interval, which is shown below:
\begin{flalign}
\label{eq:dens_WPS}
P^W_{\mathrm{tot},\bar\delta_V}(k)=\langle P^W_{\mathrm{tot},V}(k)\rangle_{\Delta\bar\delta_V},&&
\end{flalign}
where the density interval $\Delta\bar\delta_V$ is specified by the list (i)-(vi) abovementioned. These results are presented in Fig.~\ref{fig:env_WPS_tng}. Obviously, at the largest scales, $P^W_{\mathrm{tot},\bar\delta_V}(k)$ in all density intervals converge to the global WPS, while as scales become smaller, discrepancies between $P^W_{\mathrm{tot}, \bar\delta_V}(k)$ in different density intervals as well as the global WPS become more pronounced. We noticed that the mean local WPS in $\bar\delta_V>3$ is the most enhanced among all density intervals, particularly at scales around $3\ h\mathrm{Mpc}^{-1}$, where are dominated by the most massive halos, pointed out by \citet{vanDaalen2015}. As the local mean density $\bar\delta_V$ gets smaller, the mean local WPS gets smaller. In $\bar\delta_V<1$, the mean local WPSs are most severely suppressed on scales of $k\sim3\ h\mathrm{Mpc}^{-1}$. Therefore, our results suggest that most massive halos tend to reside in the densest environments, while those less dense environments show a deficit of massive halos, which are qualitatively consistent with the results of \citet{Fisher2018} and \citet{Zhu2022}.

Note that the total matter density field is formed via
\begin{flalign}
\bar\rho_\mathrm{tot}\delta_\mathrm{tot}(\mathbf{r})=\bar\rho_\mathrm{dm}\delta_\mathrm{dm}(\mathbf{r})+\bar\rho_\mathrm{gas}\delta_\mathrm{gas}(\mathbf{r})+\bar\rho_\star\delta_\star(\mathbf{r}). &&
\end{flalign}
Combing this equation and equation \eqref{eq:localWPS_3D}, the local WPS of the total matter in a sub-volume can be separated into
\begin{flalign}
\bar\rho_\mathrm{tot}^2P^W_{\mathrm{tot},V}(k) = & \bar\rho_\mathrm{dm}^2P^W_{\mathrm{dm},V}(k) + 2\bar\rho_\mathrm{dm} \bar\rho_\mathrm{gas} P^W_{\mathrm{dm-gas},V}(k) \nonumber \\
 &+ \bar\rho_\mathrm{gas}^2P^W_{\mathrm{gas},V}(k) + 2\bar\rho_\mathrm{gas}
 \bar\rho_\star P^W_{\mathrm{gas-\star},V}(k) \nonumber\\
 &+ \bar\rho_\star^2P^W_{\star,V}(k) +  2\bar\rho_\mathrm{dm} \bar\rho_\star P^W_{\mathrm{dm-\star},V}(k). &&
\label{eq:dm_gas_star_wps_wcc}
\end{flalign}
in which the cross terms $P^W_{f-g,V}(k)$ encode coherence between different matter species. This can be examined by the local WCC $C^W_{f-g,V}(k)$, which is defined as equation \eqref{eq:localWCC_3D}. The mean values of local WCCs in different density intervals $C^W_{f-g,\bar\delta_V}(k)=\langle C^W_{f-g,V}(k)\rangle_{\Delta\bar\delta_V}$ are shown in Fig. \ref{fig:env_WCC_tng}. On large scales, $C^W_{f-g,\bar\delta_V}(k)$ is close to one for all pairs of matter components in all density environments, whereas deviates more from one on smaller scales, and also shows more distinct discrepancies between different density environments. Specifically, in $\bar\delta_V>-0.5$, the correlation between the dark matter and stars remains the highest than other pairs on scales of $k\gtrsim 3 \ h\mathrm{Mpc}^{-1}$. However, this correlation decreases with decreasing density. This could be explained by the fact that stars are mostly formed in center of halos, which are very few in less dense environments. For the pair of the dark matter and gas, the correlation between them is very close to one and hardly varies with the density environment on the scales of $k\lesssim 2 \ h\mathrm{Mpc}^{-1}$. On scales smaller than $2 \ h\mathrm{Mpc}^{-1}$, this correlation first decreases with decreasing density until reaching the minimum in $-0.5<\bar\delta_V<0$ and then increases with decreasing density, so that it becomes higher even than that between the dark matter and stars in $\bar\delta_V<-0.75$. For the pair of the gas and stars, the correlation between them shows a density dependence similar to that between the dark matter and gas, but is still lower than other matter pairs in all density environments. Within the scale range we investigate, the variation of gas-dark matter and gas-stars correlations with scale and density may be related to the feedback processes, mainly active galactic nuclei (AGN) feedback at $z=0$, which can expel gas from halos to large distances and hence reshape the matter distribution.

In order to investigate the impact of the AGN feedback on the matter clustering in different density environments, we consider the ratio between the local WPS from the TNG100-1 and that from the corresponding dark matter (DM)-only simulation TNG100-1-Dark, i.e. $P^W_{f,V}(k)/P^W_{\mathrm{DM-only},V}(k)$, where $f$ refers to the total matter, dark matter and total baryons (gas+stars), respectively. The mean values of  $P^W_{f,V}(k)/ P^W_{\mathrm{DM-only}, V}(k)$ in different density intervals are plotted in Fig. \ref{fig:env_rWPS_tng}. For the total matter, we see that its global WPS is reduced on scales $k \gtrsim 1\ h\mathrm{Mpc}^{-1}$ relative to the DM-only's global WPS, with a $14\%$ suppression at $k \sim 20\ h\mathrm{Mpc}^{-1}$. This effect is also seen in the local WPS, but with some variation in different density environments. The change of the total matter clustering compared to that in the Dark-matter only simulation comes from two aspects: (1) the redistribution of baryons by non-gravitational physics, and (2) the change in the dark matter distribution resulted from the gravitational
coupling of baryons and dark matter, which is called the back-reaction \citep{vanDaalen2020}. We see from Fig. \ref{fig:env_rWPS_tng} that the former effect is the determinant of the change in the clustering of total matter for all density intervals. In particular, the clustering strength of baryons is least suppressed in the density ranges of $\bar\delta_V>3$ and $\bar\delta_V<-0.75$, revealing that the AGN feedback is the weakest in the densest and under-densest environments. Considering the fact that most massive halos prefer to reside in the densest environments, our results are consistent with the findings of \citet{Springel2018}, showing that AGN feedback has minimal impact on the most massive halos. As for the lowest AGN feedback in the under-densest environments, it may be related that these environments contain few or no halos.

%------------------------------------------------------------------------------
\section{Discussion and Conclusions}
\label{sec:concl}

The clustering of matter is an intricate process. In addition to redshift and scale dependence, the density environment plays a very important role on the clustering. When the usual Fourier statistics are utilized to examine the scale dependence, we are unable to take into account the effect of environment, since the local density information contained in the physical space is smeared out. Therefore, we need better tools to consider both scale and environment dependence simultaneously, and the continuous wavelet transform is such a tool of good performance.

In this work, we introduce some statistical quantities formulated on the specific wavelet function -- Gaussian-derived wavelet, which is a real valued continuous wavelet designed by taking the first derivative of the Gaussian smoothing function with respect to the scale parameter. With such a wavelet construction scheme, the original signal can be recovered from wavelet coefficients easily. These wavelet statistics we used include the wavelet power spectrum, the wavelet cross-correlation and the wavelet bicoherence. 

To reveal the usefulness of wavelet statistics in analyzing matter clustering, we analyze the time evolution of the 1D density distributions obtained from the Zel'dovich approximation by measuring their wavelet power spectra and bicoherences. Measurements show that the global wavelet power spectrum on small scales increases more significantly with time than that on large scales, which is generally in agreement with the Fourier case.

For manifesting the capability of wavelet statistics to perform local spectral analysis, we divide the 1D Zel'dovich density field at each time into four consecutive segments. All the local wavelet power spectra for these segments almost converge to the global wavelet power spectrum at linear stages. However, the difference between the local power spectra and the global wavelet power spectrum becomes progressively larger with time due to nonlinear effects. In particular, the wavelet power spectrum of Segment III (highly overdense environment) is significantly greater than those of other segments on scales $k\gtrsim40$, where statistical errors are smaller, which implies that structures on these scales are generated in this region at nonlinear stages. Another striking feature is that the growth of the wavelet power spectrum in Segment II (underdense environment) is severely suppressed on scales $k\gtrsim40$, meaning that there are very few structures on these scales generated in this region at later times. Moreover, measurements of the wavelet bicoherences show that the scale coupling occurs mainly in the Segment III, whereas there is almost no scale coupling in Segment II at late epochs. Both the wavelet power spectrum and the wavelet bicoherence in Segment I and IV exhibit similar behaviors to and fall between those in Segment II and III, probably because Segment I and IV are slightly overdense environments. These results demonstrate the great potential of the wavelet statistics for taking into account the effects of both environment and scale on matter clustering.

Furthermore, we generalize the 1D CWT methods to the 3D isotropic version and apply them to the 3D density fields of the TNG100-1 simulation at redshift $z=0$. Due to space limitations, wavelet bicoherences of the 3D density fields are left for future works. In order to examine the impact of the large-scale environment on the matter clustering, we split the whole simulation volume evenly to 512 small cubes and then measure the local WPS and WCC for different matter species in each sub-volume. To clearly see the dependence of the matter clustering on the density environment, we consider the mean of local WPSs and WCCs over all sub-volumes within each density interval. Our main findings are summarized as follows:
\begin{enumerate}[label=(\roman*),align=left,leftmargin=*]
	\item On the largest scales, the clustering of the total matter shows no dependence on the density environment, whereas it increases monotonically with increasing density on scales up to several times $0.1 \ h\mathrm{Mpc}^{-1}$. Compared to the global WPS of the total matter, its mean local WPS is most enhanced in the density of $\bar\delta_V>3$ around scales of $3\ h\mathrm{Mpc}^{-1}$ where are dominated by the most massive halos. As the density $\bar\delta_V$ becomes smaller and smaller, the mean local WPS is more and more suppressed around scales of $3\ h\mathrm{Mpc}^{-1}$, suggesting that lower density environments contain less massive halos. 
	\item For the pairs of the dark matter-gas, dark matter-stars and gas-stars, the cross-correlations in all density environments are close to one on large scales, whereas deviates from one significantly on small scales. Specifically, we find that the dark matter and stars are less correlated in less dense environments on small scales. The correlation between the dark matter and gas remains close to one in all environments on the scales of $k \lesssim 3\ h\mathrm{Mpc}^{-1}$. On smaller scales, this correlation first decreases and then increases with the increasing density. The correlation between the stars and gas also shows a similar density dependence, but is lower than other pairs.
	\item As revealed by the ratio between the local WPS from the TNG100-1 and that from its companion dark matter-only simulation, the impact of the AGN feedback on the matter clustering also varies with the density environment. Particularly, AGN feedback has less impact on the matter clustering in the densest environments $\bar\delta_V>3$ and the less-densest environments $\bar\delta_V<-0.75$. Compared to these extreme environments, the matter clustering strength at small scales is more suppressed by the AGN feedback in densities of $-0.75<\bar\delta_V<3$.
\end{enumerate}

These results are consistent qualitatively with previous researches about the matter clustering \citep[e.g.][]{vanDaalen2015,Fisher2018,Springel2018,vanDaalen2020,Zhu2022}. This encourages us that we can further apply statistics based on the continuous wavelet transform to various cosmological simulations and characterize the environmental dependence of the matter clustering and the baryonic effects on it in more detail.

% ------------------------------------------------------------------------------------------------
\section*{Acknowledgments}

YW especially thanks Dr. van Milligen for helpful discussions on the statistical error estimation for wavelet statistical quantities. PH acknowledges the support by the Natural Science Foundation of Jilin Province, China (No. 20180101228JC), and by the National Science Foundation of China (No. 12047569, 12147217). In this work, we used the data from IllustrisTNG simulations. The IllustrisTNG simulations were undertaken with compute time awarded by the Gauss Centre for Supercomputing (GCS) under GCS Large-Scale Projects GCS-ILLU and GCS-DWAR on the GCS share of the supercomputer Hazel Hen at the High Performance Computing Center Stuttgart (HLRS), as well as on the machines of the Max Planck Computing and Data Facility (MPCDF) in Garching, Germany.

% ------------------------------------------------------------------------------------------------
%\section*{Data availability}
%The data used in this paper are available from the correspondence author upon reasonable request.

% Appendix
\appendix
\restartappendixnumbering
%------------------------------------------------------------------------------
\section{Choice of the wavelet}

\begin{figure}[t]
	\centerline{\includegraphics[width=0.45\textwidth]{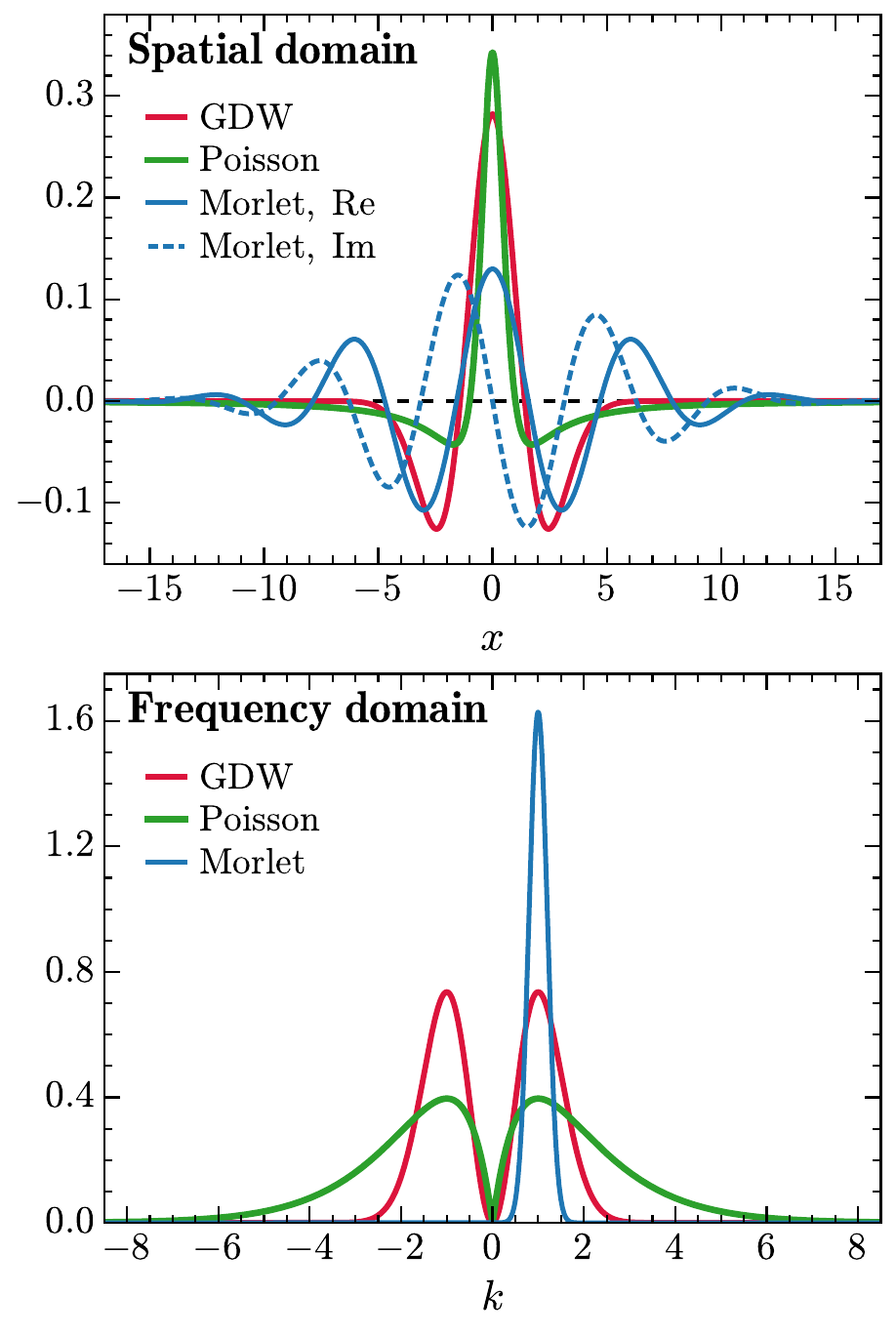}}
	\caption{Top panel: the Gaussian-derived wavelet, the Poisson wavelet and the Morlet wavelet. Bottom panel: the respective Fourier transforms of the wavelets in the top panel.}
	\label{fig:multi_wavelets}
\end{figure}

\begin{figure}[th]
	\centerline{\includegraphics[width=0.5\textwidth]{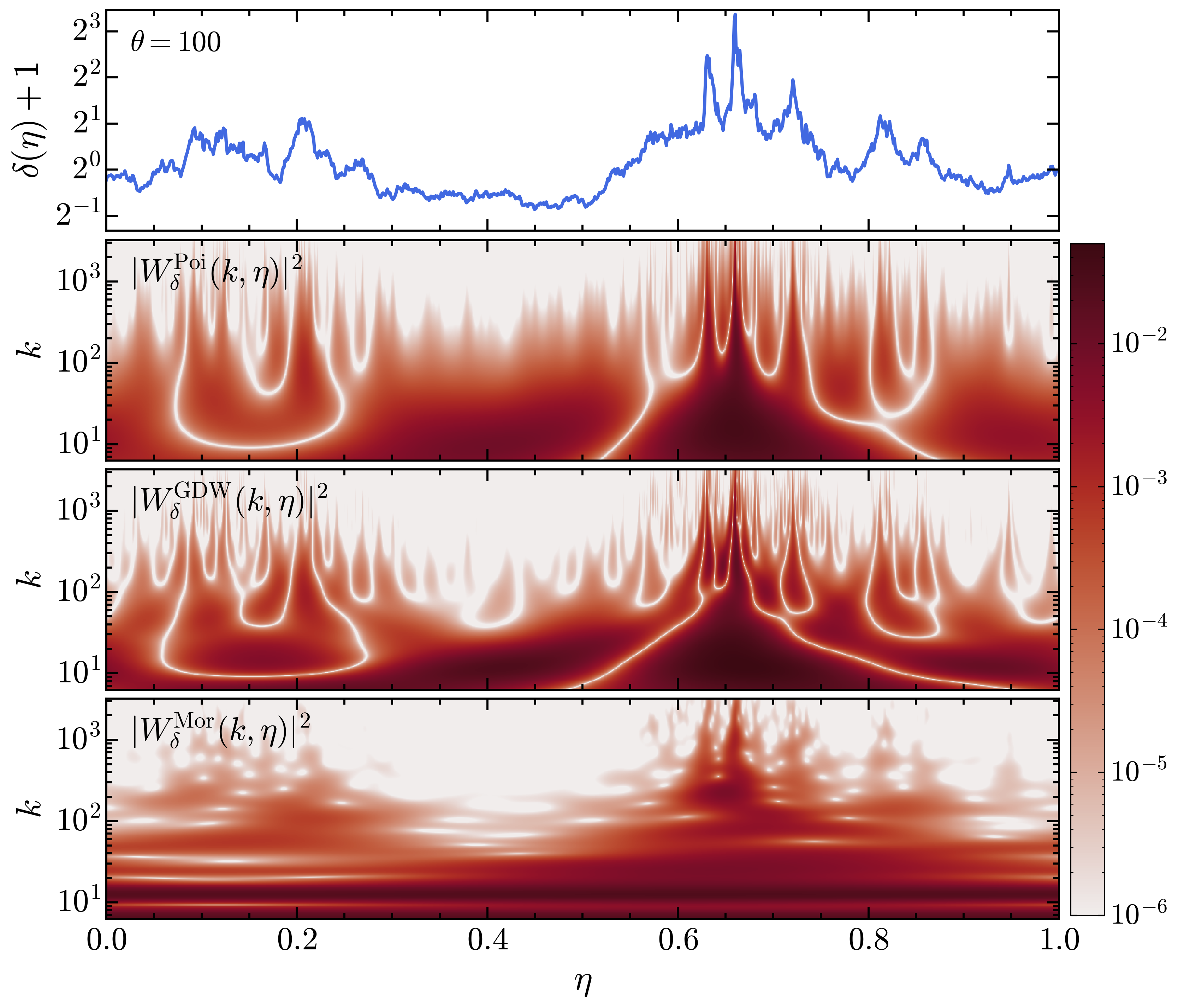}}
	\caption{First panel: the 1D density field at $\theta=100$ obtained through equation \eqref{eq:zel_dens}. \ \ Second panel: the squared wavelet transform $|W_\delta^\mathrm{Poi}(k,\eta)|^2$ of the density field based on the Poisson wavelet, where $W_\delta^\mathrm{Poi}(k,\eta)=W_\delta^\mathrm{Poi}(3w/2,\eta)=\int \delta(u) \psi^{\mathrm{Poi}}(w,x-u)\mathrm{d}u$. Here the scaled Poisson wavelet is defined as $\psi^{\mathrm{Poi}}(w,x) = \sqrt{w}\psi^{\mathrm{Poi}}(wx)$, and the relation $k=3w/2$ is obtained following the method of Section \ref{sec:scale&wavenum}. \ \ Third panel: same as the second panel, but using the GDW. \ \ Fourth panel: same as the second panel, but using the Morlet wavelet. For this wavelet, the relationship between scale parameter and the corresponding Fourier wavenumber is $k=(5+3\sqrt{3})w/10$.}
	\label{fig:multi_scalograms}
\end{figure}

In principle, there is a diverse set of wavelets available in the CWT. For instance, the most common ones are the Poisson wavelet, the Mexican hat wavelet, the spline wavelet, the Meyer wavelet, and the Morlet wavelet (see e.g. \citet{Akujuobi2022} for a review of wavelets). The choice of an appropriate wavelet depends on the goal of the performed analysis. In this study, we chose to use the GDW for two reasons. First, its shape matches well that of the density peak. Second, it provides a good trade-off between spatial and frequency resolution. Subsequently, we will highlight the superiority of the GDW by comparing it with the Poisson and the Morlet wavelets.

The Poisson wavelet described by \citep{Holschneider2000} is
\begin{flalign}
\psi^\mathrm{Poi}(x)=C^\mathrm{Poi}\frac{1-x^2}{(1+x^2)^2} &&
\label{eq:Poisson_wavelet}
\end{flalign}
with Fourier transform
\begin{flalign}
\hat\psi^\mathrm{Poi}(k)=C^\mathrm{Poi}|k|e^{-|k|},&&
\label{eq:Poisson_wavelet_k}
\end{flalign}
where the constant $C^\mathrm{Poi}$ is used to make it have the same energy as the GDW. As can be seen from Fig.~\ref{fig:multi_wavelets}, the Poisson wavelet is more localized in spatial domain but less localized in frequency domain than the GDW. The Morlet wavelet (also called Gabor wavelet) given by \citet{Hong2004} is written as
\begin{flalign}
\psi^\mathrm{Mor}(x) = C^\mathrm{Mor}e^{-x^2/(2\sigma^2)}e^{i\eta x}, &&
\label{eq:Morlet_wavelet}
\end{flalign}
which is a harmonic function with frequency $\eta$ modulated by a Gaussian envelope with variance $\sigma^2$, and its Fourier Transform is
\begin{flalign}
\hat\psi^{\mathrm{Mor}}(k) = \sqrt{2\pi}C^\mathrm{Mor}\sigma e^{-\frac{1}{2}\sigma^2(\eta+k)^2}, &&
\end{flalign}
where the constant $C^\mathrm{Mor}$ is used to make it have the same energy as the GDW. In principle, the Morlet wavelet is not a true wavelet function, since it does not satisfy the admissibility condition, i.e. $\int_0^\infty |\hat\psi^{\mathrm{Mor}}(k)|^2/k\mathrm{d}k$ is infinite. This means that the Morlet wavelet transform has no inverse transform \citep{Daubechies1992}. Only when $|\sigma \eta|$ is large enough, the admissibility condition is approximately satisfied. A reasonable choice would be $|\sigma\eta| \geq 5$ \citep{Hong2004}, and here we set $\sigma=5$, $\eta=-1$. In this case compared to the GDW and the Poisson wavelet, the Morlet wavelet is the most extended in spatial domain but the narrowest in frequency domain, as shown in Fig.~\ref{fig:multi_wavelets}.

In order to discern the applicability of the discussed wavelets, we use them individually to calculate the squared CWT of the 1D Zel'dovich density field at $\theta=100$, which is shown in Fig.~\ref{fig:multi_scalograms}. We observe that in the Poisson case, structures of the density field are resolved quite well along the spatial axis, whereas very poor along the scale axis. The Morlet wavelet yields the best separation of scales but the worst separation of spatial features among these wavelets. The GDW performance is intermediate between the Poisson and the Morlet wavelets, giving moderate spatial and scale resolution. Therefore, using GDW to perform local spectral analysis may give more reliable results.

%--------------------------------------------------------------------------------
\section{The three-dimensional isotropic GDW and CWT}
\label{sec:iso_CWT}

In general, the 3D CWT uses a scale vector containing three scale parameters, i.e. $\mathbf{w}=(w_x, w_y, w_z)$, which measures the scale along each of the three axes ($x$, $y$ and $z$), as pointed by \citet{Wang2021}. For simplicity, we consider the isotropic case, in which the same scale parameter $w$ is used for all directions. Then the isotropic CWT of the 3D field is written as
\begin{flalign}
\label{eq:isoCWT}
W_f(w,\mathbf{r}) = \int_{\mathbb{R}^3} f(\mathbf{u})\Psi(w,\mathbf{r}-\mathbf{u})\mathrm{d}^3\mathbf{u}, &&
\end{flalign}
where $\mathbb{R}^3$ represents the 3D real space, and $\Psi(w,\mathbf{r})$ is the isotropic GDW which is defined by
\begin{flalign}
\label{eq:isoGDW}
\Psi(w,\mathbf{r}) &\equiv w^\kappa\frac{\partial G(w,\mathbf{r})}{\partial w} \nonumber\\
& = \frac{w^{\kappa+2}}{16\pi^{3/2}}(6-w^2|\mathbf{r}|^2)\exp\left(-\frac{w^2|\mathbf{r}|^2}{4}\right), &&
\end{flalign}
where $G(w,\mathbf{r})=w^3\exp(-w^2|\mathbf{r}|^2/4)/(2\sqrt{\pi})^3$ is the isotropic Gaussian smoothing function, and the index $\kappa$ is set to be $-1/2$ so that the integral $\int |\Psi(w,\mathbf{r})|^2 \mathrm{d}^3\mathbf{r}$ is constant. The Fourier transform of the isotropic GDW is
\begin{flalign}
\label{eq:iso_GDW_k}
\hat\Psi(w,\mathbf{k})=2w^{\kappa-1}\frac{|\mathbf{k}|^2}{w^2}\exp\left(-\frac{|\mathbf{k}|^2}{w^2}\right). &&
\end{flalign}
In line with the method used to obtain equation \eqref{eq:inverse_cwt}, combining equations \eqref{eq:isoCWT} and \eqref{eq:isoGDW} yields the inverse CWT,
\begin{flalign}
\label{eq:inverse_isoCWT}
f(\mathbf{r}) = C+\int_0^{+\infty} w^{-\kappa}W_f(w,\mathbf{r})\mathrm{d}w.&&
\end{flalign}

Recalling that the 1D wavelet-based statistics are defined through the integral over a length $L$, the 3D wavelet-based statistics can be defined through the integral over a volume $V=L^3$. The local WPS defined in this way is shown below
\begin{flalign*}
P^{W}_{f,V}(w) = \frac{1}{V}\int_V|W_f(w,\mathbf{r})|^2\mathrm{d}^3\mathbf{r},&&
\end{flalign*}
which is the volume-average of squared wavelet coefficients within the local volume $V$, and hence can be written in a more compact form as
\begin{flalign}
\label{eq:localWPS_3D}
P^{W}_{f,V}(w) = \langle|W_f(w,\mathbf{r})|^2\rangle_V.&&
\end{flalign}
Then the local WCC between two fields $f(\mathbf{r})$ and $g(\mathbf{r})$ is given by
\begin{flalign}
\label{eq:localWCC_3D}
C^W_{f-g,V}(w) = \frac{ P^{W}_{f-g,V}(w) }{\sqrt{P^{W}_{f,V}(w)P^{W}_{g,V}(w)}},&&
\end{flalign}
where $P^{W}_{f-g,V}(w)\equiv\langle W_f(w,\mathbf{r})W_g(w,\mathbf{r})\rangle_V$.

For simulated cosmic fields with periodic box of volume $V_b=L_b^3$, the global WPS is obtained by replacing the local volume $V$ with the total volume $V_b$,
\begin{flalign}
\label{eq:globalWPS_3D}
P^{W}_{f}(w) = \langle|W_f(w,\mathbf{r})|^2\rangle_{V_b},&&
\end{flalign}
and the global WCC is
\begin{flalign}
\label{eq:globalWCC_3D}
C^W_{f-g}(w) = \frac{ P^{W}_{f-g}(w) }{\sqrt{P^{W}_{f}(w)P^{W}_{g}(w)}},&&
\end{flalign}
where $P^{W}_{f-g}(w)\equiv\langle W_f(w,\mathbf{r})W_g(w,\mathbf{r})\rangle_{V_b}$. In the 1D case, the global WPS is a smoothed version of the Fourier power spectrum, as stated by Section \ref{sec:wps&wc}. Here we go further and give a tighter relationship between the wavelet and the Fourier power spectra. According to the Parseval's theorem, we have
\begin{flalign}
	P_f^W(w) &= \langle|W_f(w,\mathbf{r})|^2 \rangle_{V_b} \nonumber\\
	&= \frac{1}{V_b}\int_{V_b} |W_f(w,\mathbf{r})|^2 \mathrm{d}^3\mathbf{r} \nonumber\\
	&= \frac{1}{(2\pi)^3V_b}\int |\hat f(\mathbf{k})|^2|\hat\Psi(w,k)|^2k^2\mathrm{d}k\sin\theta\mathrm{d}\theta\mathrm{d}\phi,
\end{flalign}
in which we exploit the fact that GDW is isotropic, namely $\hat\Psi(w,k)=\hat\Psi(w,\mathbf{k})$. If $\hat f(\mathbf{k})$ is also isotropic, then the above equation has a more simple form,
\begin{flalign}
P_f^W(w) &= \frac{1}{2\pi^2}\int_0^{+\infty} P^F_f(k)|\hat\Psi(w,k)|^2k^2\mathrm{d}k &&
\end{flalign}
where $P^F_f(k)=\frac{|\hat f(k)|^2}{V}$ is the Fourier power spectrum. Integrating the scale factor $w$ at both ends of this equation yields
\begin{flalign}
\int_0^{\infty} P_f^W(w)\mathrm{d}w &= \frac{1}{2\pi^2}\int P^F_f(k)\left(\int_0^{\infty}|\hat\Psi(w,k)|^2\mathrm{d}w\right)k^2\mathrm{d}k\nonumber\\
&=\frac{1}{4\pi^2}\int_0^{+\infty}P_f^F(k)\mathrm{d}k,&&
\end{flalign}
in which $\int_0^{\infty}|\hat\Psi(w,k)|^2\mathrm{d}w=1/2k^2$. Finally, with the help of the relation $w=2k/\sqrt{7}$ obtained by following the method of Section \ref{sec:scale&wavenum}, we find that the global wavelet and Fourier power spectra differ by only one constant factor, as shown below
\begin{flalign}
\label{eq:relation_WPS_FPS}
\frac{8\pi^2}{\sqrt{7}}P_f^W(k)=P_f^F(k).&&
\end{flalign}
% ------------------------------------------------------------------------------------------------


\begin{thebibliography}{}

\bibitem[Abbas \& Sheth(2005)]{Abbas2005} Abbas, U., \& Sheth, R. K. 2005, \mnras, 364, 1327

\bibitem[Abdulazeez et al.(2020)]{Abdulazeez2020} Abdulazeez, A. M., Zeebaree, D. Q., Zebari, D. A. et al. 2020, JSCDM, 1, 31

\bibitem[Addison(2017)]{Addison2017} Addison, P. S. 2017, The Illustrated Wavelet Transform Handbook:Introductory Theory and Applications in Science Engineering, Medicine and Finance, (2nd ed.; CRC press)

\bibitem[Addison(2018)]{Addison2018} Addison, P. S. 2018, Phil. Trans. R. Soc. A, 376, 20170258

\bibitem[Aguiar‐Conraria \& Soares(2014)]{Aguiar2014} Aguiar‐Conraria, L., \& Soares, M. J. 2014, J. Econ. Surv., 28, 344

\bibitem[Akujuobi(2022)]{Akujuobi2022} Akujuobi, C.M. 2022, Wavelets and Wavelet Transform Systems and Their Applications, (Springer, Cham)

\bibitem[Arnalte-Mur et al.(2012)]{Arnalte-Mur2012} Arnalte-Mur, P., Labatie, A., Clerc, N., et al. 2012, \aap, 542, A34

\bibitem[Cay{\'o}n et al.(2001)]{Cayon2001} Cay{\'o}n, L., Sanz, J. L., Mart{\'i}nez-Gonz{\'a}lez, E., et al. 2001, \mnras, 326, 1243

\bibitem[Chisari et al.(2019)]{Chisari2019} Chisari, N. E., Mead, A. J., Joudaki, S., et al. 2019, Open J. Astrophys., 2, 4

\bibitem[Chui(1997)]{Chui1997} Chui, C. K. 1997, Wavelets: a mathematical tool for signal analysis, (SIAM)

\bibitem[Curto et al.(2011)]{Curto2011} Curto, A., Mart{\'i}nez-Gonz{\'a}lez, E., \& Barreiro, R.B. 2011, \mnras, 412, 1038

\bibitem[Daubechies(1992)]{Daubechies1992} Daubechies, I. 1992, Ten lectures on wavelets, (SIAM)

\bibitem[Delprat et al.(1992)]{Delprat1992} Delprat, N., Escudié, B., Guillemain, P., et al. 1992, IEEE Trans. Inf. Theory, 38, 644

\bibitem[Fang \& Feng(2000)]{Fang2000} Fang, L.-Z., \& Feng, L.-L. 2000, \apj, 539, 5

\bibitem[Fisher \& Faltenbacher(2018)]{Fisher2018} Fisher, J.D., \& Faltenbacher, A. 2018, \mnras, 473, 3941

\bibitem[Flin \& Krywult(2006)]{Flin2006} Flin, P., \& Krywult, J. 2006, \aap, 450, 9

\bibitem[Frick et al.(2001)]{Frick2001} Frick, P., Beck, R., Berkhuijsen, E. M., et al. 2001, \mnras, 327, 1145

\bibitem[Fujiwara \& Soda(1996)]{Fujiwara1996} Fujiwara, Y., \& Soda, J. 1996, Prog. Theor. Phys., 95, 1059

\bibitem[Gabor(1946)]{Gabor1946} Gabor, D. 1946, J.IEEE, 93, 429

\bibitem[Gao \& Yan(2011)]{Gao2011} Gao, R.X., \& Yan, R. 2011, in Wavelets, (Springer, Boston, MA), 17

\bibitem[Gonz{\'a}lez-Nuevo et al.(2006)]{Gonzalez2006} Gonz{\'a}lez-Nuevo, J., Arg{\"u}eso, F., L{\'o}pez-Caniego, M., et al. 2006, \mnras, 369, 1603

\bibitem[Gouda \& Nakamura(1989)]{Gouda1989} Gouda, N., \& Nakamura, T. 1989, Prog. Theor. Phys, 81, 633

\bibitem[Gu et al.(2013)]{Gu2013} Gu, J.-H., Xu, H.-G., Wang J.-Y., et al. 2013, \apj, 773, 38

\bibitem[He et al.(2006)]{He2006} He, P., Liu, J., Feng, L.-L. 2006, \prl, 96, 051302

\bibitem[Holschneider(2000)]{Holschneider2000} Holschneider, M. 2000, in Wavelets in the Geosciences, ed. R. Klees, R. Haagmans, (Springer, Berlin, Heidelberg), 1

\bibitem[Hong \& Kim(2004)]{Hong2004} Hong, J.C., \& Kim, Y.Y. 2004, Exp. Mech., 44, 387

\bibitem[Hudgins et al.(1993)]{Hudgins1993} Hudgins, L., Friehe, C. A., \& Mayer, M. E. 1993, \prl, 71, 3279

\bibitem[Kaiser \& Hudgins(1994)]{Kaiser1994} Kaiser, G., \& Hudgins, L. H. 1994, A friendly guide to wavelets, (Boston: Birkhäuser)

\bibitem[Khalifa et al.(2008)]{Khaliffa2008} Khalifa, O. O., Harding, S. H., \& Hashim, A. H. A. 2008, SPIJ, 2, 17

\bibitem[Labatie et al.(2012)]{Labatie2012} Labatie, A., Starck, J. L., \& Lachi\`{e}ze-Rey, M. 2012, \apj, 746, 172

\bibitem[Liu \& Fang(2008)]{Liu2008} Liu, J.-R., \& Fang, L.-Z. 2008, \apj, 672, 11

\bibitem[Lu et al.(2010)]{Lu2010} Lu, Y., Zhu, W.-S., Chu, Y.-Q., et al. 2010, \mnras, 408, 452

\bibitem[Mallat(2009)]{Mallat2009} Mallat, S. 2009, A Wavelet Tour of Signal Processing: The Sparse Way, (3rd ed.; Academic Press)

\bibitem[Man et al.(2019)]{Man2019} Man, Z.-Y., Peng, Y.-J., Kong, X. et al. 2019, \mnras, 488, 89

\bibitem[Manfredi et al.(2016)]{Manfredi2016} Manfredi, G., Rouet, J. L., Miller, B., et al. 2016, \pre, 93, 042211

\bibitem[Marinacci et al.(2018)]{Marinacci2018} Marinacci, F., Vogelsberger, M., Pakmor, R., et al. 2018, \mnras, 480, 5113

\bibitem[Mart\'{i}nez et al.(1993)]{Martinez1993} Mart{\'i}nez, V. J., Paredes, S., \& Saar, E. 1993, \mnras, 260, 365

\bibitem[Meyers et al.(1993)]{Meyers1993} Meyers, S. D., Kelly, B. G., \& O'Brien, J. J. 1993, Mon. Wea. Rev., 121, 2858

\bibitem[Miller \& Rouet(2010)]{Miller2010} Miller, B. N., \& Rouet, J. L. 2010, J. Stat. Mech. Theory Exp., P12028.

\bibitem[Naiman et al.(2018)]{Naiman2018} Naiman, J. P., Pillepich, A., Springel, V., et al. 2018, \mnras, 477, 1206

\bibitem[Nelson et al.(2018)]{Nelson2018} Nelson, D., Pillepich, A., Springel, V., et al. 2018, \mnras, 475, 624

\bibitem[Pando \& Fang(1996)]{Pando1996} Pando, J., \& Fang, L.-Z. 1996, \apj, 459, 1

\bibitem[Pando et al.(1998)]{Pando1998} Pando, J., Lipa, P., Greiner, M. et al. 1998, \apj, 496, 9

\bibitem[Pando et al.(2004)]{Pando2004} Pando, J., Feng, L.-L., \& Fang, L.-Z. 2004, \apj, 154, 475

\bibitem[Peng et al.(2010)]{Peng2010} Peng, Y.-J. et al. 2010, \apj, 721, 193

\bibitem[Pillepich et al.(2018)]{Pillepich2018} Pillepich, A., Nelson, D., Hernquist, L., et al. 2018, \mnras, 475, 678

\bibitem[Roh et al.(2019)]{Roh2019} Roh, S., Ryu, D., \& Kang, H. 2019, \apj, 883, 138

\bibitem[Romeo et al.(2004)]{Romeo2004} Romeo, A.B., Horellou, C., \& Bergh, J. 2004, \mnras, 354, 1208

\bibitem[Rozgacheva et al.(2012)]{Rozgacheva2012} Rozgacheva, I. K., Boriso, A. A., Agapov, A. A., et al. 2012, preprint (arXiv:1201.5554)

\bibitem[Schwinn et al.(2018)]{Schwinn2018} Schwinn, J., Baugh, C. M., Jauzac, M., et al. 2018, \mnras, 481, 4300

\bibitem[Shandarin \& Zeldovich(1989)]{Shandarin1989} Shandarin, S. F., \& Zel'dovich, Y. B. 1989, RMP, 61, 185.

\bibitem[Shi et al.(2018)]{Shi2018} Shi, X., Nagai, D., \& Lau, E. T. 2018, \mnras, 481, 1075

\bibitem[Soda \& Suto(1992)]{Soda1992} Soda, J., \& Suto, Y. 1992, \apj, 396, 379

\bibitem[Springel(2010)]{Springel2010} Springel, V. 2010, \mnras, 401, 791

\bibitem[Springel et al.(2018)]{Springel2018} Springel, V., Pakmor, R., Pillepich, A., et al. 2018, \mnras, 475, 676

\bibitem[Starck et al.(2004)]{Starck2004}  Starck, J.L., Aghanim, N., \& Forni, O. 2004, \aap, 416, 9

\bibitem[Tabatabaei et al.(2013)]{Tabatabaei2013} Tabatabaei, F. S., Berkhuijsen, E. M., Frick, P., et al. 2013, \aap, 557, A129

\bibitem[Tatekawa \& Maeda(2001)]{Tatekawa2001} Tatekawa, T., \& Maeda, K. I. 2001, \apj, 547, 531

\bibitem[Tegmark et al.(2004)]{Tegmark2004} Tegmark M. et al. 2004, \apj, 606, 702

\bibitem[Tian et al.(2011)]{Tian2011} Tian, H.-J., Neyrinck, M. C., Budav\'{a}ri, T., et al. 2011, \apj, 728, 34

\bibitem[Torrence \& Compo(1998)]{Torrence1998} Torrence, C., \& Compo, G. P. 1998, Bull. Am. Meteorol. Soc., 79, 61

\bibitem[van Daalen \& Schaye(2015)]{vanDaalen2015} van Daalen, M. P., \& Schaye, J. 2015, \mnras, 452, 2247

\bibitem[van Daalen et al.(2020)]{vanDaalen2020} van Daalen, M. P., McCarthy, I. G., Schaye, J. 2020, \mnras, 491, 2424

\bibitem[Van den Berg(2004)]{Van den Berg2004} Van den Berg, J. C. 2004, Wavelets in physics, (Cambridge University Press)

\bibitem[van Milligen et al.(1995a)]{Milligen1995a} van Milligen, B. P., Hidalgo, C., \& S\'{a}nchez, E. 1995, \prl, 74, 395

\bibitem[van Milligen et al.(1995b)]{Milligen1995b} van Milligen, B. P., S\'{a}nchez, E., Estrada, T., et al. 1995, Phys. Plasmas, 2, 3017

\bibitem[van Milligen et al.(1997)]{Milligen1997} van Milligen, B. P., Hidalgo, C., S\'{a}nchez, E., et al. 1997, Rev. Sci. Instrum., 68, 967

\bibitem[Wang et al.(2018)]{Wang2018} Wang, Y. et al. 2018, \apj, 868, 130

\bibitem[Wang \& He(2021)]{Wang2021} Wang, Y., \& He P. 2021, Commun. Theor. Phys., 73, 095402

\bibitem[Yang et al.(2020)]{Yang2020} Yang, H.-Y., He, P., Zhu, W.-S. et al. 2020, \mnras, 498, 4411

\bibitem[Zel'dovich(1970)]{Zeldovich1970} Zel'dovich, Y. B. 1970, \aap, 5, 84

\bibitem[Zhu et al.(2022)]{Zhu2022} Zhu, W.-S., Zhang, F.-P., \& Feng, L.-L. 2022, \apj, 924, 132

\end{thebibliography}
\end{document}